\documentclass[aps, prb, twocolumn, showpacs, 10pt, floatfix, superscriptaddress, floatfix, longbibliography]{revtex4-2}
\makeatletter
\newcommand{\hideappendixsubsections}{%
  \let\l@subsection\@gobbletwo
}
\makeatother

\usepackage{lipsum}  
\usepackage[english]{babel}
\usepackage[utf8]{inputenc}
\usepackage[T1]{fontenc}

\usepackage{amsmath}
\usepackage{braket}
\usepackage{amssymb}
\usepackage[normalem]{ulem}
\usepackage{dsfont}
\usepackage{bm}
\usepackage{bbm}
\usepackage[dvipsnames]{xcolor}
\usepackage[x11names]{xcolor}
\usepackage{stmaryrd}
\usepackage{bbm}
\usepackage{mathtools}
\usepackage{hyperref}
\usepackage{physics}

\usepackage{tikz,pgfplots}
\usetikzlibrary{plotmarks}

\usepackage{graphicx}
\usepackage[export]{adjustbox} 
\usepackage{dcolumn}
\usepackage{bm}
\usepackage{xcolor}
\usepackage{hyperref}
\hypersetup{
    colorlinks=true,
    citecolor=Green,
    linkcolor=Blue,
    urlcolor=Blue
    }

\makeatletter
\newcommand{\thickhline}{%
\noalign {\ifnum 0=`}\fi \hrule height 1.2pt
\futurelet \reserved@a \@xhline%
}
\newcolumntype{"}{@{%
\hskip\tabcolsep\vrule width 1.2pt\hskip\tabcolsep}%
}
\makeatother

\renewcommand{\Im}{\operatorname{Im}}

\newcommand{\ii}{\mathrm{i}}
\newcommand{\de}{\mathrm{d}}
\newcommand{\ve}{\varepsilon}

\newcommand{\sfT}{\mathsf{T} \, }
\newcommand{\sfaT}{\tilde{\mathsf{T}}}

\pgfplotsset{compat=1.18}
\begin{document}


\preprint{APS/123-QED}

\title{Dissipation--enhanced scrambling in the SYK model coupled to a lossy cavity}

\author{Pietro Pelliconi}
\email{pelliconi@princeton.edu}
\affiliation{%
Department of Physics, Princeton University, Princeton, New Jersey 08544, USA%
}%

\author{Bastien Lapierre}
\email{bastien.lapierre@phys.ens.fr}
\affiliation{%
Philippe Meyer Institute, Physics Department, École Normale Supérieure (ENS), Université PSL, 24 rue Lhomond, F-75231 Paris, France%
}%

\author{Shinsei Ryu}
\email{shinseir@princeton.edu}
\affiliation{%
Department of Physics, Princeton University, Princeton, New Jersey 08544, USA%
}%

\date{\today}


\begin{abstract}
    We study the Yukawa-Sachdev-Ye-Kitaev model, a disordered model of $N$ Majorana fermions and $R=\gamma N$ bosons in which the bosons are linearly coupled to independent realizations of SYK $p$--body interactions, in the presence of dissipation, modeled by a Lindblad master equation. Motivated by recent proposals for implementing SYK models in quantum simulators, we focus on bosonic leakage at rate $\kappa$. Initializing the system in the steady state, we analyze the late-time fermionic relaxation rate and the Lyapunov exponent, solving the large-$N$ theory both numerically and for $p$ large, finding a rich landscape of dynamical behaviors. Most notably, the Lyapunov exponent remains positive for every value of $\kappa$ and, for $p>2$, can even grow as $\kappa$ increases. The QED case $p=2$, which lies between the fully chaotic regime $p>2$ and the integrable case $p=1$, exhibits special features. We also identify a critical value of the boson-to-fermion ratio $\gamma_c \approx 2/p^2$ separating distinct dynamical regimes.
\end{abstract}


\maketitle

\tableofcontents

\section{Introduction}
\label{sec:Introdution}

Many body quantum chaos encompasses a broad range of phenomena in isolated quantum systems, including spectral correlations, eigenstate thermalization, and the dynamical spreading of information~\cite{DAlessio2016QuantumChaos}. Among these phenomena, quantum information scrambling provides a particularly sharp characterization of how quantum systems lose memory of their initial conditions: initially simple operators evolve into increasingly complex many-body operators, thereby delocalizing initially local information~\cite{Yasuhiro_Sekino_2008, Aleiner2016Butterfly, Nahum2018OperatorSpreading, vonKeyserlingk2018OperatorHydrodynamics, doi:10.1126/science.abg5029}.
Scrambling can be diagnosed through out-of-time-ordered correlators (OTOCs), whose growth is exponential in systems possessing a parametrically separated scrambling regime~\cite{Larkin1969Quasiclassical,Shenker2014Butterfly,Hosur2016Chaos,Patel2017Butterfly,Roberts2018OperatorGrowth, Swingle_Nature_2018}. The corresponding growth rate, the quantum Lyapunov exponent, has therefore become a central diagnostic of chaotic many-body dynamics~\cite{Maldacena2016Bound}. Among the few interacting models in which this quantity can be computed analytically, the Sachdev--Ye--Kitaev (SYK) model~\cite{sachdev1993, sachdev2015, Kitaev2015SimpleModelPart1, Kitaev2015SimpleModelPart2} occupies a central position, combining solvability in the large--$N$ limit with rapid---and, in its low-temperature regime, maximal---scrambling~\cite{Kitaev2015SimpleModelPart1, Kitaev2015SimpleModelPart2, Maldacena:2016hyu}.

Realistic quantum systems, however, are inevitably coupled to their environment, making dissipative many-body dynamics both experimentally relevant and theoretically challenging~\cite{Daley2014QuantumTrajectories, Sieberer2016Keldysh}. Non-unitary evolution restricts the analytical tools available, and controlled solutions of dissipative quantum chaos remain scarce. Recent works on dissipative dynamics in SYK models, based on the Lindblad equation, have provided important first steps toward analytically tractable descriptions of open many-body chaos, revealing that dissipation can strongly modify and, in many situations, suppress scrambling~\cite{PhysRevB.106.075138, PhysRevResearch.4.L022068, Bhattacharjee_2024, PhysRevB.108.075110, PhysRevD.107.106006, ozaki2026exactspectrumanomalousrelaxation, Chen:2017dbb, Bhattacharjee2023OperatorGrowth, Garcia-Garcia:2024tbd, Liu:2024stj}.

These results fit into a broader body of work pointing toward the general expectation that coupling to an environment suppresses operator growth and, at sufficiently strong dissipation, can destroy exponential scrambling altogether~\cite{Chen:2017dbb, Schuster:2022bot, PhysRevLett.131.220404, PhysRevLett.130.250401, Bhattacharjee2023OperatorGrowth, Garcia-Garcia:2024tbd, Liu:2024stj}. This picture becomes less straightforward, however, when the dissipated degrees of freedom are themselves responsible for mediating the interactions. Such a situation arises naturally in cavity and circuit quantum electrodynamics, where photons or phonons generate effective long-range interactions between matter degrees of freedom while simultaneously leaking out of the experimental apparatus~\cite{Ritsch2013Cavity, Mivehvar2021CavityQED, Blais2021CircuitQED}. Increasing the leakage rate can both weaken the coherent interaction and modify the fluctuations of the mediating field, giving rise to a potentially richer dependence of scrambling on dissipation. Nevertheless, an analytically tractable dissipative many-body model exhibiting this physics has yet to be identified.

\begin{figure}[t]
    \centering
    \includegraphics[width=\columnwidth]{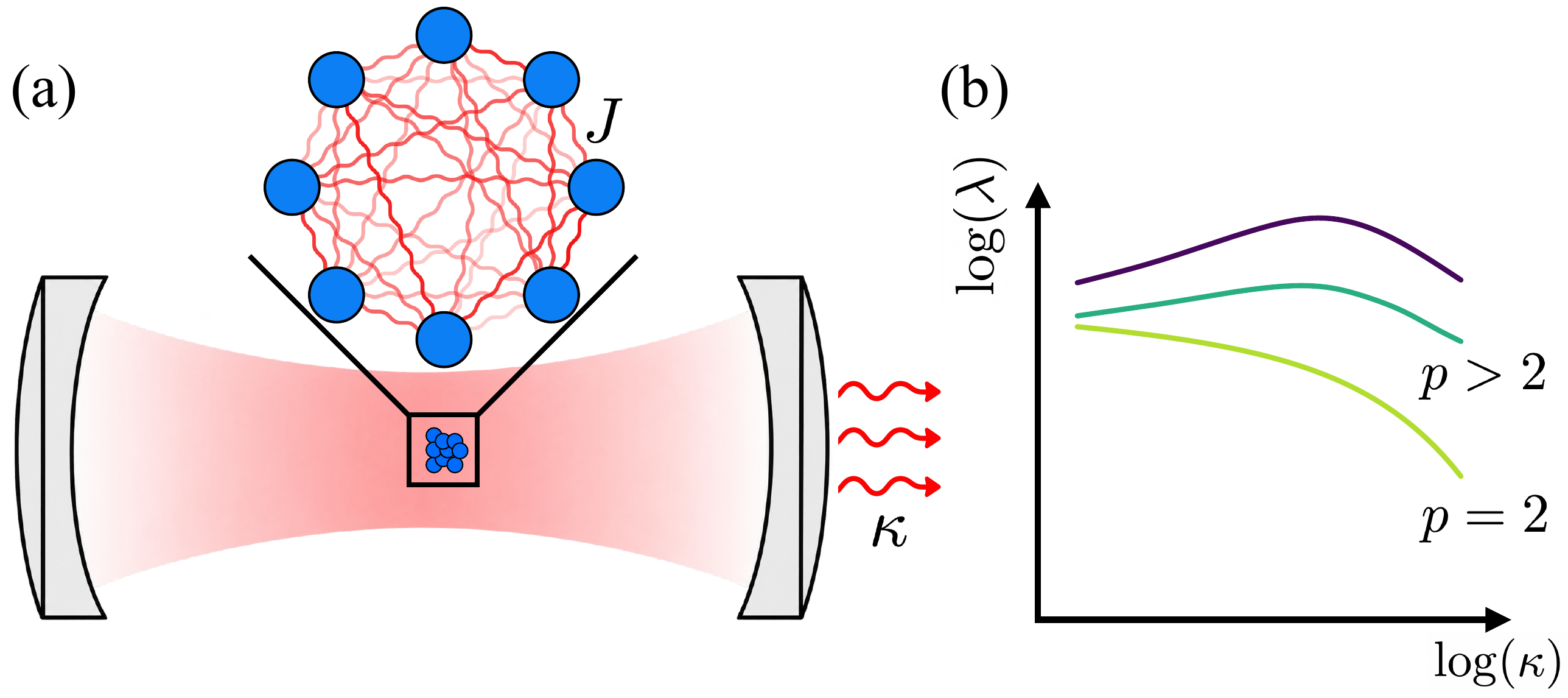}
    \caption{ We study the dissipative dynamics of the $p$--body YSYK model in the presence of bosonic leakage $\kappa$. (a) Sketch of a possible realization in cQED platforms, where bosonic interactions are mediated by dispersive bosons, which can also leak out of the cavity. (b) Sketch of the dissipation--enhanced scrambling: the Lyapunov exponent $\lambda$ stays positive for any $p$ and any leakage strength $\kappa$, and is enhanced at finite $\kappa$ for $p>2$.}
    \label{fig:setup1}
\end{figure}

In this work, we bridge this gap by analyzing the effect of photon loss in the Yukawa--SYK (YSYK) model~\cite{Marcus:2018tsr,Kim:2019lwh,PhysRevB.100.115132,PhysRevLett.124.017002, Wang:2020dtj}, as illustrated in Fig.~\ref{fig:setup1}(a). The model consists of fermions (Majorana in our case \footnote{We employ Majorana fermions primarily for analytical convenience. Although number-conserving complex fermions would more closely mirror the experimental implementations cited above, at half-filling we expect the qualitative mechanisms governing relaxation and scrambling to be closely related to those studied here}) coupled to bosonic modes through independent random $p$-body interactions, and has recently been proposed as a promising route toward realizing SYK physics in cavity quantum electrodynamics (cQED)~\cite{uhrich2023cavityquantumelectrodynamicsimplementation, baumgartner2025quantumsimulationsachdevyekitaevmodel, baumgartner2025, wntd-53rd, Bode:2026wvb} (see also~\cite{chen2018, creffield2026}).
Within cQED platforms, photon leakage is unavoidable, making it natural to ask how such dissipation affects information scrambling. From the theoretical perspective, photon dissipation in the YSYK model remains amenable to analytical treatment in the large-$p$ limit using the Lindblad master equation, allowing us to compute both relaxation rates and Lyapunov exponents. We find a rich, non-monotonic behavior of both quantities as functions of the dissipation rate and the boson--to--fermion ratio.

\textit{Summary of results} --- Initializing the system in the steady state of the dissipative Lindbladian dynamics, we study the late--time fermionic relaxation rate in Sec.~\ref{sec:Late-time_relaxation}, and the Lyapunov exponent in~\ref{sec:Lyapunov-growth}, both analytically at large--$p$ in the strict auxiliary boson limit and numerically without resorting to such approximations.

The relaxation rate $\Gamma$, which characterizes the decay of two-point correlators, is governed by two energy scales: an effective interaction strength $J_{\rm eff}$ and a dissipative rate $\Gamma_{\rm Pur}$ associated with the {\it Purcell effect}. At weak photon loss $\kappa$, the system exhibits anomalous relaxation, whereas for larger $\kappa$ and boson--to--fermion ratio $\gamma>2/p^2$, the relaxation rate becomes nonmonotonic, initially increasing with $\kappa$ before being suppressed at stronger dissipation.

The Lyapunov exponent $\lambda$, extracted from the late-time growth of out-of-time-order correlators, is controlled by the same energy scales $J_{\rm eff}$ and $\Gamma_{\rm Pur}$, but exhibits a richer dependence on $p$, $\gamma$ and $\kappa$. In contrast to previous studies~\cite{Bhattacharjee_2024, Bhattacharjee2023OperatorGrowth, Garcia-Garcia:2024tbd, Liu:2024stj}, we find that the Lyapunov exponent remains positive for any value of the dissipation strength. For small $\kappa$, it is generically suppressed, and for $p = 2$ it decreases monotonically, eventually vanishing universally as $\kappa^{-3}$. 
The case $p > 2$ displays qualitatively different behavior: for $\gamma > 2/p^2$, there exists an intermediate dissipation regime where the Lyapunov exponent is \textit{enhanced} by quantum fluctuations of the bath, before eventually vanishing as $\kappa^{-1}$, as schematically illustrated in Fig.~\ref{fig:setup1}(b). Conversely, for $\gamma < 2/p^2$, $\lambda$ decreases monotonically.

Our results demonstrate that dissipation does not necessarily eliminate scrambling: the Lyapunov exponent can remain positive and, in some regimes, can even be enhanced by quantum fluctuations of the dissipative bosonic sector. We further find that $p=2$ displays distinct behavior, interpolating between the integrable $p=1$ case and the more strongly chaotic $p>2$ YSYK models. Finally, we identify $\gamma_c=2/p^2$ as the critical boson--to--fermion ratio separating qualitatively different dynamical regimes.

The remainder of the paper is organized as follows. In Sec.~\ref{sec:Setup}, we introduce the dissipative Yukawa-SYK model and formulate its Lindbladian dynamics within the Schwinger-Keldysh (SK) path-integral framework. In Sec.~\ref{sec:Late-time_relaxation}, we determine the late-time relaxation rate using both exact large-$p$ solutions and numerical calculations. In Sec.~\ref{sec:Lyapunov-growth}, we carry out an analogous analysis of the Lyapunov exponent. We conclude in Sec.~\ref{sec:Conclusion} with a discussion of possible extensions and future directions. Technical details are provided in the Appendices.

\section{Setup}
\label{sec:Setup}

\subsection{Yukawa SYK in the dispersive limit}
\label{sec:YSYK_setup}

We consider a Majorana YSYK model in which $R$ bosons are linearly coupled to $N$ Majorana fermions through disordered SYK$_p$--like interactions~\cite{Marcus:2018tsr, Kim:2019lwh, PhysRevB.100.115132}. This model has attracted considerable attention in recent years, both because it reproduces key features of the SYK model, most notably maximal scrambling in the strongly coupled low-energy regime, suggesting the possibility of a holographic dual, and because of promising proposals for experimental realizations~\cite{uhrich2023cavityquantumelectrodynamicsimplementation, baumgartner2025quantumsimulationsachdevyekitaevmodel, baumgartner2025, wntd-53rd, Bode:2026wvb, chen2018, creffield2026}.

We introduce $R$ bosonic coordinates $x_{\mu}$ and canonically conjugated momenta $\pi_\mu$ satisfying $[x_\mu, \pi_\nu] = \ii \, \delta_{\mu \nu}$, together with $N$ Majorana fermions $\psi^i$ obeying the anti-commutation relations $\{ \psi^i, \psi^j \} = \delta^{ij}$, which imply $(\psi^i)^2 = \frac{1}{2}$. Throughout the paper, Greek indices denote bosonic coordinates, while Roman indices label fermionic degrees of freedom.

The Hamiltonian of YSYK reads
\begin{multline}
    H = \frac{1}{2} \sum_{\mu = 1}^R \Big( (\pi^{\mu})^2 + \Delta^2 (x^{\mu})^2 \Big) \\
    + \ii^{\frac{p}{2}} \sqrt{2\Delta} \sum_{\mu=1}^R \sum_{i_{1}\dots i_p}^N J_{i_1 \dots i_p}^{\mu} x^{\mu} \psi^{i_1} \dots \psi^{i_p} 
    \label{eq:YSYK_Hamiltonian}
\end{multline}
where the Fermionic subscripts are always ordered as $\{i_1 < \dots < i_p\}$, and the coupling constants are drawn from independent Gaussian distributions with zero mean and variance
\begin{equation}
    \mathbb E \big[ J_{i_1 \dots i_p}^2 \big] = \frac{(p-1)!}{N^p} \, J^2 \ .
    \label{eq:variance_of_disordered_couplings}
\end{equation}
The factor of $\sqrt{\Delta}$ in the interaction vertex ensures that the parameter $J$ has the dimension of an energy, setting the energy scales of the system. The order of fermionic interaction $p$ is a generic positive even integer, in order to respect spin-statistics. By convention, Majorana operators satisfy $\big(\psi^i\big)^\dagger = \psi^i$, so that the phase $\ii^{\frac{p}{2}}$ in~\eqref{eq:YSYK_Hamiltonian} makes the Hamiltonian a Hermitian operator.

In order to obtain an extensive free energy, the number of bosonic modes is taken to scale linearly with the number of fermions at large-$N$, namely
\begin{equation}
    R = \gamma N \ .
    \label{eq:gamma_definition}
\end{equation}

\textit{The auxiliary limit} --- In the most general setting, the dynamics of the Yukawa--SYK model can be highly nontrivial, depending on the relative strength of the couplings and on the initial state considered. This complexity is familiar, for instance, from the phenomenology of the Caldeira--Leggett model, where memory effects and revivals can arise.

An interesting simplifying regime, which we consider in the present work, is the limit in which the bosonic degrees of freedom become auxiliary to the fermionic ones, namely $\Delta \gg J$. In this regime, the bosons can be integrated out, yielding an effective low-rank SYK$_{2p}$ interaction~\cite{Marcus:2018tsr, Kim:2019lwh}. Here, however, we retain the bosonic degrees of freedom explicitly, as their dissipative dynamics is the central focus of this work.

It is then useful to rewrite the Hamiltonian~\eqref{eq:YSYK_Hamiltonian} in terms of bosonic ladder operators, defined as
\begin{align}
    a_{\mu} \, = & \; \sqrt{\frac{\Delta}{2}} \,  \Big( x^\mu + \frac{\ii}{\Delta} \, \pi^{\mu}  \Big) \ ,\\
    a_{\mu}^{\dagger} \, = & \; \sqrt{\frac{\Delta}{2}} \,  \Big( x^\mu - \frac{\ii}{\Delta} \, \pi^{\mu}  \Big) \ ,
\end{align}
with canonical commutation relations $\big[ a_{\mu}, a_{\nu}^\dagger \big] = \delta_{\mu \nu}$. In these coordinates, the Hamiltonian~\eqref{eq:YSYK_Hamiltonian} becomes 
\begin{multline}
    H = \sum_{\mu = 1}^R \Delta \, a^\dagger_{\mu} a_{\mu} \\
    + \ii^{\frac{p}{2}} \sum_{\mu=1}^R \sum_{i_{1}\dots i_p}^N J_{i_1 \dots i_p}^{\mu} \big( a^\dagger_{\mu} + a_{\mu} \big) \psi^{i_1} \, \dots \psi^{i_p} \ .
    \label{eq:YSYK_Hamiltonian_ladder_operators}
\end{multline}
While in principle one could consider a different detuning $\Delta_{\mu}$ for each bosonic mode, as done previously in {\it e.g.},~\cite{Kim:2019lwh,chen2018, uhrich2023cavityquantumelectrodynamicsimplementation}, here we take $\Delta_{\mu} \equiv \Delta$ for all modes $\mu$. This choice is natural for the platforms considered in~\cite{baumgartner2025quantumsimulationsachdevyekitaevmodel, baumgartner2025}.

\subsection{Bosonic dissipation}
\label{sec:YSYK_dissipation}

In this work, we study Markovian dissipation described by a Lindblad master equation acting on the full boson--fermion system. Denoting by $\rho$ the joint density matrix of bosons and fermions, we consider the dynamics
\begin{align}
    \partial_t \rho \, \equiv & \; \mathcal L (\rho) \nonumber \\
    \, = & \; - \ii \big[ H, \rho \big] + \sum_{\alpha} \Big( L_\alpha \rho L_\alpha^\dagger - \frac{1}{2} \big\{ L_\alpha^\dagger L_\alpha, \rho \big\} \Big) \ ,
    \label{eq:Lindbladian_dyanamics_generic_master_equation}
\end{align}
where the indices $\alpha$ label the different jump operators $L_{\alpha}$.
A natural dissipative process in this setting is bosonic leakage at rate $\kappa$, described by the Lindblad jump operators
\begin{equation}
    L_{\mu} = \sqrt{\kappa} \, a_{\mu} \ .
    \label{eq:Dissipative_jump_operator}
\end{equation}
Physically, such a jump operator removes bosonic excitations from the system over time.

Our interest in this dissipative dynamics is motivated both experimentally and theoretically. On the experimental side, the YSYK model has attracted attention as a promising route toward realizing SYK physics in quantum simulators. A central challenge in such implementations is understanding how unavoidable dissipation affects the targeted chaotic dynamics. In particular, the jump operator~\eqref{eq:Dissipative_jump_operator} captures a ubiquitous loss mechanism in quantum simulators where photons or phonons mediate interactions between matter degrees of freedom. Indeed, excitations of the mediating bosonic modes can escape the experimental apparatus, and~\eqref{eq:Dissipative_jump_operator} provides a natural effective description of such processes.

The same dissipative process is also interesting from a theoretical perspective. As evident from the Hamiltonian~\eqref{eq:YSYK_Hamiltonian}, the effective fermionic interactions arise from virtual photonic exchange, which generates the disordered many-body dynamics responsible for chaotic behavior. At the same time, photon leakage reduces the lifetime of the mediating degrees of freedom and modifies the fluctuations of the bosonic sector. Understanding the competition between these two effects is therefore a natural question. As we will show, this competition leads to a non-trivial dependence of chaotic observables, such as the relaxation rate and the Lyapunov exponent, on the dissipation strength.

\subsection{Schwinger-Keldysh path integral}
\label{sec:YSYK_SK_collective_field}

We initialize the system at $t=0$ in the steady state $\rho_{\rm ss}$ of the dissipative dynamics. In Appendix~\ref{app:steady_state}, we show that, in a perturbative expansion around the auxiliary boson limit (namely in powers of $\Delta^{-1}$), this steady state takes the form
\begin{equation}
    \rho_{\rm ss} = \ketbra{0}  \otimes  \frac{1}{2^{N/2}} \, \mathbbm 1  + \mathcal O \big( \Delta^{-1} \big) 
    \label{eq:Initial_density_matrix}
\end{equation}
where the bosonic modes are approximately in their vacuum state, while the fermions are at infinite temperature. The first non-trivial correction to this expression is also computed in Appendix~\ref{app:steady_state}.

To study correlation functions in this state, we employ a real-time path-integral formulation. In particular, we consider the real-time partition function
\begin{equation}
    Z = \Tr \Big[ e^{t \mathcal L} \big( \rho_{\rm ss} \big) \Big] \equiv 1 \ ,
    \label{eq:Keldysh_partition_function}
\end{equation}
where the equality on the RHS follows from the fact that the Lindbladian master equation~\eqref{eq:Lindbladian_dyanamics_generic_master_equation} generates a completely positive and trace--preserving evolution. The steady-state condition can equivalently be implemented in the path-integral formulation by extending the initial time to the distant past, $t_i\to-\infty$, and selecting time-translation invariant solutions.

The path-integral approach is particularly convenient in the present context, since it allows us to formulate the theory directly in the large--$N$ limit and to derive closed equations for the collective degrees of freedom. In Appendix~\ref{app:path-integrals} we study in detail the path integral representation of~\eqref{eq:Keldysh_partition_function}. As usual in disordered systems, it is convenient to consider the averaged partition function, which is best expressed in terms of the collective field variables 
\begin{align}
    G_{\sigma, \sigma'}(t,t') \, = & \;  - \frac{\ii}{N} \sum_{i = 1}^N \psi^i_\sigma(t) \psi^i_{\sigma'}(t') \ , \\
    D_{\sigma, \sigma'}(t,t') \, = & \;  - \frac{\ii}{R} \sum_{\mu = 1}^R x_\sigma^\mu(t) x_{\sigma'}^{\mu}(t') \ ,
\end{align}
for the Majorana fermions and for the bosons, respectively. The indices $\{\sigma, \sigma'\} \in \{\pm, \pm\}$ represent the branch in the contour for the operator insertion, and, when the two are in the same branch, there is an additional time--ordering involved. We refer the Reader to Appendix~\ref{app:SK_collective_fields} for a detailed explanation. Let us also remark that, since we are in the steady state, all two--point function depend only on the time difference.

As is customary in the Schwinger--Keldysh formalism, we define $G_{\mathsf{T}}(t) \equiv G_{++}(t)$ and $G_{\tilde{\mathsf{T}}}(t) \equiv G_{--}(t)$, while the {\it lesser} and {\it greater} functions are $G_{<}(t) \equiv G_{+-}(t)$ and $G_{>}(t) \equiv G_{-+}(t)$, respectively. These quantities can equivalently be expressed in terms of the {\it retarded}, {\it advanced} and {\it Keldysh} correlators as
\begin{align}
    G_{\mathsf T}(t) \, = & \; \frac{1}{2} \big( G_{\rm K}(t) + G_{\rm A}(t) + G_{\rm R}(t) \big) \ , \label{eq:RAK_to_time_1}  \\
    G_{<}(t) \, = & \; \frac{1}{2} \big( G_{\rm K}(t) + G_{\rm A}(t) - G_{\rm R}(t) \big) \ , \label{eq:RAK_to_time_2}  \\
    G_{>}(t) \, = & \; \frac{1}{2} \big( G_{\rm K}(t) - G_{\rm A}(t) + G_{\rm R}(t) \big) \ , \label{eq:RAK_to_time_3} \\
    G_{\tilde{\mathsf{T}}}(t) \, = & \; \frac{1}{2} \big( G_{\rm K}(t) - G_{\rm A}(t) - G_{\rm R}(t) \big) \ ,\label{eq:RAK_to_time_4}
\end{align}
where evidently 
\begin{equation}
    G_{\mathsf T}(t) + G_{\tilde{\mathsf{T}}}(t) = G_{<}(t) + G_{>}(t) \ .
\end{equation}
The same definitions apply to the bosonic collective fields.

The collective variables can be introduced in the effective partition function through Lagrange multipliers $\Sigma_{ab}(t)$ and $\Pi_{ab}(t)$, which will play the role of fermionic and bosonic self--energies, respectively. The resulting effective partition function takes the form
\begin{equation}
    Z_{\rm eff} = \int \prod_{\sigma \sigma' = \pm} \! \! \mathcal D \big[ G \Sigma D \Pi \big]_{\sigma \sigma'} \, \, e^{\ii S_{\rm eff} [ G \Sigma D \Pi ]} \ ,
\end{equation}
with 
\begin{widetext}
\begin{multline}
    \frac{\ii}{N} \, S_{\rm eff} \big[ G \Sigma D \Pi \big] = \frac{1}{2} \Tr \log \Big( \mathbf G_0^{-1}(\omega) - \mathbf \Sigma(\omega) \Big) - \frac{\gamma}{2} \Tr \log \Big( \mathbf D_0^{-1}(\omega) - \mathbf \Pi(\omega) \Big) \\
    + \sum_{\sigma, \sigma'} \frac{1}{2} \, \eta_{\sigma} \eta_{\sigma'} \int \de t \, \de t' \, \Sigma_{\sigma' \sigma}(t',t)  G_{\sigma \sigma'}(t,t') - \frac{\gamma}{2} \, \eta_{\sigma} \eta_{\sigma'} \int \de t \, \de t' \, \Pi_{\sigma' \sigma}(t',t)  D_{\sigma \sigma'}(t,t') \\
    +(-\ii)^{p+1} \frac{\gamma J^2 \Delta}{p} \sum_{\sigma,\sigma'} \int \de t \, \de t' \eta_{\sigma} \eta_{\sigma'} D_{\sigma \sigma'}(t,t') G^p_{\sigma \sigma'}(t,t') \ ,
    \label{eq:Bilocal_effective_action_main_text}
\end{multline}
\end{widetext}
where the signs $\eta_{\pm} = \pm 1$ implement the appropriate SK contour ordering, and the boldface notation indicates that the collective fields are understood as matrices in contour space. We refer to Appendix~\ref{app:SK_collective_fields} for further details.

The overall factor of $N$ in the effective action allows us to study the theory at large--$N$ in the saddle-point approximation. Varying~\eqref{eq:Bilocal_effective_action_main_text} with respect to the self-energies yields the Schwinger--Dyson (SD) equations \footnote{Our convention for the Fourier transform is
\begin{equation*}
    f(\omega) = \int \de t \, f(t) \, e^{\ii \omega t}
\end{equation*}}
\begin{align}
    G_{\rm R}(\omega) \, = & \; \; \frac{1}{\omega + \ii \varepsilon - \Sigma_{\rm R}(\omega)} \ , \label{eq:GR_SD_eq} \\
    G_{\rm A}(\omega) \, = & \; \; \frac{1}{\omega - \ii \varepsilon - \Sigma_{\rm A}(\omega)} \ , \\
    G_{\rm K}(\omega) \, = & \; G_{\rm R} (\omega) \Sigma_{\rm K}(\omega) G_{\rm A}(\omega) \ , \textcolor{white}{\Big)} \label{eq:GK_SD_eq}
\end{align}
for the fermionic correlators, and
\begin{align}
    D_{\rm R}(\omega) \, = & \; \frac{1}{\big( \omega + \frac{\ii \kappa}{2} \big)^2 - \Delta^2 - \Pi_{\rm R}(\omega)} \ , \label{eq:DR_SD_eq} \\ 
    D_{\rm A}(\omega) \, = & \; \frac{1}{\big( \omega - \frac{\ii \kappa}{2} \big)^2 - \Delta^2 - \Pi_{\rm A}(\omega)} \ , \label{eq:DA_SD_eq} \\
    D_{\rm K}(\omega) \, = & \; D_{\rm R} (\omega) \Big( \Pi_{\rm K}(\omega) - \big[ D_{\rm K}^0 \big]^{-1} \Big) D_{\rm A}(\omega) \ , \label{eq:DK_SD_eq}
\end{align}
for the bosonic ones. These equations make explicit the interpretation of $\Sigma$ and $\Pi$ as fermionic and bosonic self-energies, respectively. Conversely, varying~\eqref{eq:Bilocal_effective_action_main_text} with respect to the collective fields gives
\begin{align}
    \Sigma_{>} (t) \, = & \; \ii^{p-1} \, 2 \Delta \gamma J^2 \, D_{>}(t) G_{>}^{p-1}(t)  \ , \label{eq:SD_Sigma_gtr} \\
    \Sigma_{<} (t) \, = & \; \ii^{p-1} \, 2 \Delta \gamma J^2 \, D_{<}(t) G_{<}^{p-1}(t) \ , \label{eq:SD_Sigma_lss}
\end{align}
and 
\begin{align}
    \Pi_{>} (t) \, = & \; - \ii^{p+1}  \, \frac{2 \Delta J^2 }{p} \, G_{>}^{p}(t)  \ ,  \label{eq:SD_Pi_gtr} \\
    \Pi_{<} (t) \, = & \; - \ii^{p+1}  \, \frac{2 \Delta J^2 }{p} \, G_{<}^{p}(t) \ . \label{eq:SD_Pi_lss}
\end{align}
The SD equations above admit a natural diagrammatic interpretation in terms of polarization loops. Representing fermionic and bosonic dressed propagators by straight and wiggly lines, respectively, the dominant large--$N$ contributions are the melonic diagrams
\begin{equation}
    \Sigma(t) \; = \;  \text{\includegraphics[width=2.9cm, valign=c]{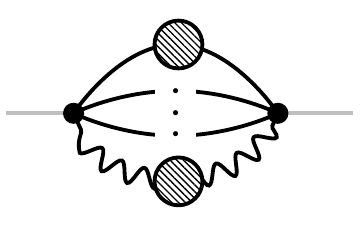}}
    \label{eq:Sigma_polarization_diagram}
\end{equation}
for the fermionic self energy, and similarly
\begin{equation}
    \Pi(t) \; = \;  \text{\includegraphics[width=2.9cm, valign=c]{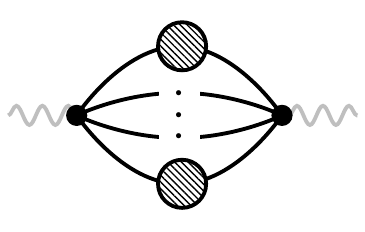}}
    \label{eq:Pi_polarization_diagram}
\end{equation}
for the bosonic one. Together, these saddle-point equations provide a closed description of the dynamics in terms of the SK collective fields.

To see explicitly the closure of the system, suppose that the solutions for $G_{\rm R/A/K}(t)$ and $D_{\rm R/A/K}(t)$ are known. The relations (\ref{eq:RAK_to_time_1}--\ref{eq:RAK_to_time_4}) then determine $G_{> / <}(t)$ and $D_{> / <}(t)$. The SD equations (\ref{eq:SD_Sigma_gtr}--\ref{eq:SD_Pi_lss}) subsequently give $\Sigma_{> / <}(t)$ and $\Pi_{> / <}(t)$. Finally, using
\begin{align}
    \Sigma_{\rm K}(t) \, = & \; \Sigma_{>}(t) + \Sigma_{<}(t) \ , \label{eq:Sigma_K_from_lss_and_gtr} \\ 
    \Sigma_{\rm R}(t) \, = & \; \theta(t) \, \Big( \Sigma_{>}(t) - \Sigma_{<}(t) \Big) \ ,  \label{eq:Sigma_R_from_lss_and_gtr} \\
    \Sigma_{\rm A}(t) \, = & \; \theta(-t) \, \Big( \Sigma_{<}(t) - \Sigma_{>}(t) \Big) \ ,  \label{eq:Sigma_A_from_lss_and_gtr}
\end{align}
for fermions and
\begin{align}
    \Pi_{\rm K}(t) \, = & \; \Pi_{>}(t) + \Pi_{<}(t) \ , \label{eq:Pi_K_from_lss_and_gtr} \\ 
    \Pi_{\rm R}(t) \, = & \; \theta(t) \, \Big( \Pi_{>}(t) - \Pi_{<}(t) \Big) \ ,  \label{eq:Pi_R_from_lss_and_gtr} \\
    \Pi_{\rm A}(t) \, = & \; \theta(-t) \, \Big( \Pi_{<}(t) - \Pi_{>}(t) \Big) \ .   \label{eq:Pi_A_from_lss_and_gtr}
\end{align}
for bosons, one obtains the Keldysh components of the self-energies. The loop is then closed by using (\ref{eq:GR_SD_eq}--\ref{eq:DK_SD_eq}), which determines the Keldysh components of the two-point functions and completes the self-consistent solution.

In the next Section, we solve these equations both analytically in the limit of large interaction order $p$, and numerically, focusing in particular on the case $p = 2$.

\section{Late-time relaxation rate}
\label{sec:Late-time_relaxation}

The first observable that we consider is the late--time fermionic relaxation rate, {\it i.e.}, the exponential rate at which the two--point function vanishes at late times. Focusing {\it e.g.} on the retarded two--point function, we generally have
\begin{equation}
    \lim_{t \to \infty} G_{\rm R}(t) =  - \ii \theta(t) \, \mathsf{C} \, e^{- \Gamma t} + o \big( e^{- \Gamma t} \big) \ ,
    \label{eq:late_time_relaxation_definition}
\end{equation}
where $\Gamma$ is defined as the relaxation rate, $\theta(t)$ is the Heaviside step--function, and $\mathsf C$ is a positive real number. For dissipative systems, it is natural to focus on such observables, as dissipation is expected to suppress correlations at late times. A similar phenomenon occurs in chaotic systems, where unitary chaotic dynamics also leads to the decay of correlations at large time separations~\cite{DAlessio2016QuantumChaos}. Moreover, in recent years, several works have investigated the interplay between these two effects, highlighting in particular the non--commutativity of the thermodynamic and weak--dissipation limits as a characteristic feature of many--body scrambling systems~\cite{PhysRevB.106.075138, PhysRevResearch.4.L022068, PhysRevD.107.106006, Tomaz_Prosen_2002, PROSEN2004244, Prosen_2007, PhysRevB.109.064311, PhysRevB.111.104303, PhysRevE.110.054204, YoshimuraSa2024, r8qy-dcgq, ozaki2026exactspectrumanomalousrelaxation}. This phenomenon, referred to as {\it anomalous relaxation}, is closely related to Ruelle–-Pollicott resonances \cite{PhysRevLett.56.405, Pollicott_1985}, which were originally introduced to characterize the decay of correlations in classical chaotic systems and have since been extended to quantum and quantum many-body dynamics \cite{Tomaz_Prosen_2002, PROSEN2004244, Prosen_2007}. 

With this motivation, it is natural in our context to investigate $\Gamma$ throughout the full parameter space, and in particular for different values of $\kappa$ and $\gamma$. We find a rich variety of behaviors. In the following sections, we derive an analytical prediction for the fermionic two--point function at large $p$ and compare it with numerical solutions. In both approaches, as we will show, extracting the late--time relaxation rate is straightforward.

\subsection{\texorpdfstring{large--$p$}{Large-p} solution}
\label{sec:Late-time_relaxation_large-p}

In this Section, we solve the Schwinger--Dyson equations in the large--$p$ limit. The analysis simplifies considerably in the auxiliary regime, where we assume $|\omega| \ll \{\Delta, \kappa\}$, as will become clear below. We start from the standard ansatz  
\begin{align}
    G_{>}(t) \, = & \; - \frac{\ii}{2} \, \exp \bigg( \frac{g_>(t)}{p} \bigg) \ , \\ 
    G_{<}(t) \, = & \; \frac{\ii}{2} \, \exp \bigg( \frac{g_<(t)}{p} \bigg) \ ,
\end{align}
where the Majorana nature of the fermions imposes the boundary conditions $g_{>}(0) = g_{<}(0) = 0$. At leading order in the auxiliary limit, the fermionic stationary state is maximally mixed (see Appendix~\ref{app:steady_state}) and factorizes from the bosonic sector, implying $[\rho_{\rm ss},\psi^i]=0$. The cyclicity of the trace then gives $G_{>}(t)=-G_{<}(t)$, which within the large--$p$ parametrization implies
\begin{equation}
    g_{>}(t) = g_{<}(t)\equiv g(t).
\end{equation}
This relation is preserved self-consistently by the Schwinger--Dyson equations in the strict auxiliary limit.
Using this solution in (\ref{eq:SD_Pi_gtr}--\ref{eq:SD_Pi_lss}), and taking $p$ to be even, we obtain
\begin{equation}
    \Pi_{>} (t) = - \ii  \, \frac{\Delta J^2 }{2^{p-1}p} \, e^{g(t)} = \Pi_{<} (t) \ .
\end{equation}
This equality implies that the retarded and advanced bosonic self-energies vanish, $\Pi_{\rm R}(t)=\Pi_{\rm A}(t)=0$, while the Keldysh component is given by $\Pi_{\rm K}(t)=2\Pi_{>}(t)$. Therefore, using (\ref{eq:DR_SD_eq}--\ref{eq:DA_SD_eq}), we obtain
\begin{equation}
    D_{\rm R}(\omega) = - \frac{1}{\Delta^2 + \frac{\kappa^2}{4}} = D_{\rm A}(\omega) \ ,
    \label{eq:large_p_retarded_and_advanced_propagators_frequency}
\end{equation}
where we have used the auxiliary boson approximation to neglect powers of $\omega$ in the free bosonic propagator. Using~\eqref{eq:DK_SD_eq}, we find
\begin{equation}
    D_{\rm K}(t) = - \frac{\ii \kappa \, \delta(t)}{\Delta \big( \Delta^2 + \frac{\kappa^2}{4} \big)} - \frac{\ii \Delta J^2 \, e^{g(t)}}{2^{p-2} p \big(\Delta^2 + \frac{\kappa^2}{4}\big)^2} \ .
    \label{eq:large-p_bosonic_Keldysh_correlator}
\end{equation}
Since $D_{\rm R}(t)=D_{\rm A}(t)$ in this limit, the Keldysh relations (\ref{eq:RAK_to_time_2}--\ref{eq:RAK_to_time_3}), now applied to bosonic correlators, give
\begin{equation}
    D_{>}(t) = - \frac{\ii \kappa \, \delta(t)}{2 \Delta \big( \Delta^2 + \frac{\kappa^2}{4} \big)} - \frac{\ii \Delta J^2 \, e^{g(t)}}{2^{p-1} p \big(\Delta^2 + \frac{\kappa^2}{4}\big)^2} \ ,
    \label{eq:D_gtr_large_p}
\end{equation}
and, in the auxiliary limit, $D_{<}(t)=D_{>}(t)$. Substituting this result into (\ref{eq:SD_Sigma_gtr}--\ref{eq:SD_Sigma_lss}) determines the fermionic greater and lesser self-energies, which can then be related to $\Sigma_{\rm R}(t)$ through~\eqref{eq:Sigma_R_from_lss_and_gtr}. The resulting system closes into an equation relating $\Sigma_{\rm R}(t)$ and $G_{\rm R}(t)$, which can be rewritten as a Liouville equation for the field $g(t)$. Expanding~\eqref{eq:GR_SD_eq} for small $\Sigma_{\rm R}(\omega)$, we obtain
\begin{equation}
    G_{\rm R}(\omega) = \frac{1}{\omega + \ii \ve} + \frac{1}{\omega^2} \, \Sigma_{\rm R}(\omega) + \dots \ ,
\end{equation}
while at the same time
\begin{equation}
    G_{\rm R}(t) = - \ii \theta(t)  - \frac{\ii \theta(t)}{p}  g(t)  + \dots \ ,
\end{equation}
which implies \footnote{In this derivation we are neglecting to resolve indeterminate expressions such as $\delta(t)\theta(t)$. We check the validity of~\eqref{eq:dissipative_Liouville_equation} by comparing it to the numerical solution of Fig.~\ref{fig:Relaxation_rate_p_2}, which is obtained without resorting to any auxiliary approximation.}
\begin{equation}
    \frac{\de^2 g(t)}{\de t^2} = - \frac{\gamma p J^2 \kappa \, \delta(t)}{2^{p-2} \big(\Delta^2 + \frac{\kappa^2}{4}\big)} - \frac{ \gamma \, \Delta^2 J^4 \, e^{2 g(t)}}{2^{2p-4} \big(\Delta^2 + \frac{\kappa^2}{4}\big)^2 }  \ .
    \label{eq:dissipative_Liouville_equation}
\end{equation}
This equation is a Liouville equation supplemented by an attractive contact interaction. Imposing the boundary condition $g(0)=0$ and the discontinuity of the first derivative induced by the contact term, we obtain the solution
\begin{equation}
    e^{g(t)} = \frac{\mathsf{A}}{J_{\rm eff}\cosh(\mathsf{A} |t| + \mathsf{B})} \ ,
    \label{eq:complete_Liouville_Solution}
\end{equation}
where the parameters in the solution are defined as
\begin{equation}
    \mathsf{A} = \sqrt{J_{\rm eff}^2 + \Gamma_{\rm Pur}^2} \ , \qquad \sinh(\mathsf{B}) =  \Gamma_{\rm Pur} / J_{\rm eff} \ ,
    \label{eq:large-p_definitions_A_B}
\end{equation}
with 
\begin{align}
    J_{\rm eff} \, = & \; \frac{\sqrt{\gamma} \, \Delta J^2}{2^{p-2} \big(\Delta^2 + \frac{\kappa^2}{4}\big) } \ , \label{eq:J_eff_definition} \\
    \Gamma_{\rm Pur} \, = & \; \frac{p \gamma J^2 \kappa}{2^{p-1} \big(\Delta^2 + \frac{\kappa^2}{4}\big)} \ . \label{eq:Gamma_Pur_definition}
\end{align}
It is interesting to note that a solution of the form~\eqref{eq:complete_Liouville_Solution} also appears in large--$p$ unitary SYK~\cite{Maldacena:2016hyu}, where instead $\mathsf B=0$. Dissipation modifies this solution by generating a cusp around $t=0$, which leads to a faster suppression of correlations at short times \footnote{This short-time suppression should not be confused with the late--time relaxation rate $\Gamma$.}. The origin of this cusp can be traced back to the sign of the contact term in the bosonic Keldysh correlator~\eqref{eq:large-p_bosonic_Keldysh_correlator}. A sign change would reverse the orientation of the cusp and would, for instance, lead to $|G_{<}(t)|>1/2$ in a finite interval close to the origin. Such behavior is incompatible with completely positive and trace--preserving dynamics generated by a Lindbladian, providing an additional consistency check of~\eqref{eq:large-p_bosonic_Keldysh_correlator}.

Using the solution~\eqref{eq:Gamma_Pur_definition}, the retarded fermionic two--point function reads
\begin{equation}
    G_{\rm R}(t) = - \ii \theta(t) \Bigg( \frac{\mathsf{A}}{J_{\rm eff}\cosh(\mathsf{A} |t| + \mathsf{B})} \Bigg)^{\! \frac{1}{p}}\ ,
    \label{eq:complete_Liouville_Solution_retarded}
\end{equation}
from which we obtain the late--time relaxation rate 
\begin{equation}
    \Gamma = \frac{1}{p} \sqrt{J_{\rm eff}^2 + \Gamma_{\rm Pur}^2} \ .
    \label{eq:Late_time_relaxation_rate}
\end{equation}
The result shows that the relaxation rate depends on the combination of two energy scales, $J_{\rm eff}$ and $\Gamma_{\rm Pur}$. Their physical origin can be understood from a simple quantum-optical example.

In quantum optics, energy scales of the form~\eqref{eq:J_eff_definition} and~\eqref{eq:Gamma_Pur_definition} naturally arise from the interplay between dispersive interactions and photon loss. As a simple example, consider a two--level qubit $\{ \ket{g}, \ket{e} \}$ dispersively coupled to a dissipative cavity, with Hamiltonian
\begin{equation}
    H = \omega_{\rm c} \, a^\dagger a + \omega_{\rm q} \, \sigma_{+} \sigma_{-} + J \Big( a^\dagger \sigma_- + \sigma_{+} a \Big) \ ,
\end{equation}
where $\sigma_{\pm} = \big( \sigma_{x} \pm \ii \sigma_{y}\big)/2$ are the raising and lowering qubit operators. The system is also subject to a Markovian bath with jump operator $L = \sqrt{\kappa} \, a$. Going to the rotating frame comoving with the qubit frequency $\omega_{\rm q}$, the Hamiltonian becomes 
\begin{equation}
    H_{\rm rf} = \Delta \, a^\dagger a + J \Big( a^\dagger \sigma_- + \sigma_{+} a\Big) \ ,
\end{equation}
where we defined the detuning $\Delta = \omega_{\rm c} - \omega_{\rm q}$. We consider the {\it bad cavity} limit, where $\{ \Delta, \kappa \} \gg J$. In this limit, the cavity adjusts to the qubit state much faster than any other timescale, so we can {\it adiabatically eliminate} it using the Lindblad master equation in the Heisenberg picture, imposing
\begin{multline}
    \dot a = i \big[ H_{\rm rf} , a \big] + L^\dagger a L - \frac{1}{2} \big\{ L^\dagger L, a \big\} \\
    = - \ii \big( \Delta a + J \sigma_-  \big) - \frac{\kappa}{2} \, a \approx 0 \ ,
    \label{eq:adiabatic_elimination_for_a}
\end{multline}
and similarly for $a^\dagger$. This implies a qubit--reduced Lindbladian master equation of the form
\begin{multline}
    \dot \rho_q  = \ii \frac{\Delta J^2}{\Delta^2 + \frac{\kappa^2}{4}} \big[ \sigma_+ \sigma_-, \rho_q \big] \\
    + \frac{\kappa J^2}{\Delta^2 + \frac{\kappa^2}{4}} \Big( \sigma_- \rho_q \sigma_+ - \frac{1}{2} \big\{ \sigma_+ \sigma_-,  \rho_q \big\} \Big) \ ,
    \label{eq:reduced_Lindblad_master_equation_adiabatic_elimination}
\end{multline}
upon substituting~\eqref{eq:adiabatic_elimination_for_a} into the full qubit--photon Lindblad master equation. From~\eqref{eq:reduced_Lindblad_master_equation_adiabatic_elimination}, we see the emergence of two energy scales similar in form to~\eqref{eq:J_eff_definition} and~\eqref{eq:Gamma_Pur_definition}, with a natural interpretation. In particular, the qubit attains a second--order virtual photon exchange that mediates the unitary interaction with an effective energy scale of the form
\begin{equation}
    J_{\rm eff} \propto \frac{\Delta J^2}{\Delta^2 + \frac{\kappa^2}{4}} \ ,
\end{equation}
as in~\eqref{eq:J_eff_definition}. On the other hand, from the Lindblad master equation~\eqref{eq:reduced_Lindblad_master_equation_adiabatic_elimination} it is clear that excited qubits de--excite at the {\it Purcell decay rate}
\begin{equation}
    \Gamma_{\rm q} = \frac{\kappa J^2}{\Delta^2 + \frac{\kappa^2}{4}} \propto \Gamma_{\rm Pur} \ .
    \label{eq:Purcell_effect_QO}
\end{equation}
Indeed, initializing a density matrix in the state $\rho_{\rm q}(0) = \ketbra{e}{e}$, and parametrizing $\rho_{\rm q}(t) = \rho_{\rm g}(t)\ketbra{g}{g} + \rho_{\rm e}(t)\ketbra{e}{e}$, it is straightforward to find that the solution to~\eqref{eq:reduced_Lindblad_master_equation_adiabatic_elimination} is
\begin{align}
    \rho_{\rm e}(t) \, = & \; e^{- \Gamma_{\rm q} t} \ , \\
    \rho_{\rm g}(t) \, = & \; 1 - e^{- \Gamma_{\rm q} t} \ .
\end{align}
This effect, with the particular functional form of~\eqref{eq:Purcell_effect_QO}, is commonly called the {\it Purcell effect} in the quantum optics literature~\cite{Purcell1946}.
All in all, the energy scales in~\eqref{eq:Late_time_relaxation_rate} are expected from the above quantum optics perspective. What is interesting in the many--body case of YSYK is that the relaxation rate depends on both $J_{\rm eff}$ and $\Gamma_{\rm Pur}$. Indeed, in the simple qubit model, the unitary part of the evolution does not show any sign of relaxation, as the system is integrable.

The SYK model, in contrast, is a chaotic many--body system in which correlations decay even at infinite temperature. Therefore, the result~\eqref{eq:Late_time_relaxation_rate} suggests a clear competition between many--body quantum--chaotic relaxation and dissipation due to the Markovian bath. 
Another notable feature is the different dependence of $J_{\rm eff}$ and $\Gamma_{\rm Pur}$ on the boson--to--fermion ratio $\gamma$, which is $\sqrt{\gamma}$ for the former and linear for the latter. The different dependence on $\gamma$ has a simple origin. For the effective interaction strength $J_{\rm eff}$, different realizations of disorder combine incoherently. Since the variance of a sum of independent random variables is additive, the effective variance scales as $J_{\rm eff}^2\to\gamma J_{\rm eff}^2$, leading to the scaling $J_{\rm eff} \propto \sqrt{\gamma}$~\cite{baumgartner2025}.
Conversely, the dissipative contribution scales linearly with the number of decay channels, as already evident from the qubit example above when multiple photonic modes are included. This explains the different $\gamma$ dependence of the two energy scales.

Let us now analyze the behavior of $\Gamma$ across the different parameter regimes. The two contributions entering~\eqref{eq:Late_time_relaxation_rate} display qualitatively different dependences on the dissipation strength $\kappa$. The effective interaction scale $J_{\rm eff}$ decreases monotonically as $\kappa$ increases, reflecting the fact that stronger photon leakage suppresses virtual photon exchange. In the large dissipation regime, $\kappa \gg \Delta$, it decays as $\kappa^{-2}$. Conversely, the dissipative scale $\Gamma_{\rm Pur}$ displays the characteristic non--monotonic behavior of the Purcell effect: it increases linearly for small $\kappa$, reaches a maximum at $\kappa = 2 \Delta$, and eventually decreases as $\kappa^{-1}$ at large dissipation.

As a consequence, for $\kappa \gg \Delta$, the Purcell contribution dominates and the relaxation rate behaves as $\Gamma \sim \kappa^{-1}$. At small and intermediate dissipation strengths, however, the competition between the two scales can lead to either monotonic or non--monotonic behavior. We now make this statement precise. Rewriting~\eqref{eq:Late_time_relaxation_rate} as
\begin{equation}
    \Gamma = \frac{J^2}{2^{p-2} p \Delta} \, \Big( 1 + \frac{\kappa^2}{4 \Delta^2} \Big)^{-1} \sqrt{\gamma + \gamma^2 p^2 \frac{\kappa^2}{4 \Delta^2}} \ ,
    \label{eq:Late_time_relaxation_rate_rewritten}
\end{equation}
which suggests introducing the dimensionless variables
\begin{equation}
    x \equiv \frac{\kappa}{2\Delta} \ , \qquad \qquad \tilde \Gamma =  \frac{2^{p-2} p \Delta}{J^2} \, \Gamma \ ,
\end{equation}
and expanding at small $x$ gives
\begin{multline}
    \tilde \Gamma(x) = \frac{1}{1+x^2} \sqrt{\gamma + \gamma^2 p^2 x^2} \\
    = \sqrt{\gamma} + \sqrt{\gamma} \Big( \frac{p^2 \gamma}{2} -1 \Big) x^2 + \dots\ .
\end{multline}
This expansion shows that the monotonicity properties of the relaxation rate are controlled entirely by the boson--to--fermion ratio $\gamma$. In particular,
\begin{equation}
    \begin{cases}
        \gamma \leq \frac{2}{p^2} & \to \; \; \Gamma \text{ monotonic} \ , \\
        \gamma > \frac{2}{p^2} & \to \; \; \Gamma \text{ non-monotonic} \ .
    \end{cases} \label{eq:monotonicity_relaxation_rate_gamma}
\end{equation}  
Thus, for $\gamma < 2/p^2$, the effective interaction scale dominates and the relaxation rate decreases monotonically with increasing $\kappa$. Conversely, for $\gamma > 2/p^2$, the contribution from $\Gamma_{\rm Pur}$ becomes sufficiently strong to induce a non--monotonic dependence. The value
\begin{equation}
    \gamma_{\rm c}=\frac{2}{p^2}
\end{equation}
therefore separates the two qualitatively distinct regimes.

\subsection{Numerical extrapolation for \texorpdfstring{$p=2$}{p=2}}
\label{sec:Late-time_relaxation_p-2}

\begin{figure}[t]
    \centering
    \includegraphics[width=\columnwidth]{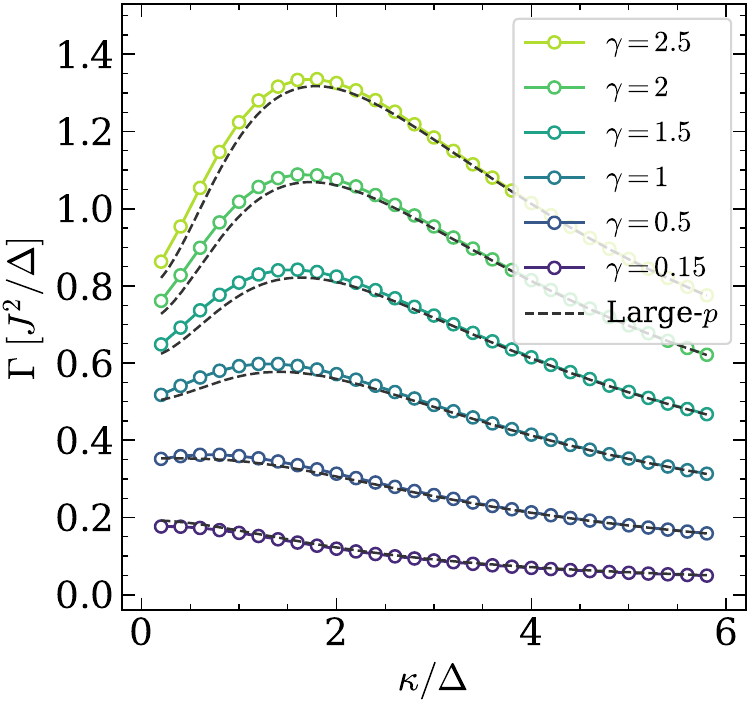}
    \caption{Late time relaxation rate $\Gamma$ for $p=2$. The dots have been obtained solving numerically the Schwinger--Dyson equations (\ref{eq:GR_SD_eq}--\ref{eq:SD_Pi_lss}) choosing $\Delta = 1$ and $J = 0.05$. For this choice of parameters the auxiliary approximation hold. We then compare the large--$p$ result~\eqref{eq:Late_time_relaxation_rate}, finding remarkable agreement.}
    \label{fig:Relaxation_rate_p_2}
\end{figure}

To benchmark the analytical prediction~\eqref{eq:Late_time_relaxation_rate} and assess its validity at finite interaction order, we determine the late--time relaxation rate numerically for $p=2$. This is done by solving the system of Schwinger--Dyson equations (\ref{eq:GR_SD_eq}--\ref{eq:SD_Pi_lss}) iteratively. We initialize the fermionic and bosonic two--point functions with their free dissipative values, $G_0(\omega)$ and $D_{0}(\omega)$, and iteratively update the self-consistent equations to obtain the new solutions $\tilde G_{n}(\omega)$ and $\tilde D_{n}(\omega)$. The fermionic propagator is then updated according to
\begin{equation}
    G_{n}(\omega) = (1 - \alpha) \, G_{n-1}(\omega) + \alpha \, \tilde G_{n}(\omega) \ ,
\end{equation}
and analogously for $D_{n}(\omega)$. The parameter $0 < \alpha \leq 1$ controls the amount of damping introduced in the iteration in order to avoid numerical instabilities; throughout our analysis we use $\alpha=0.12$. The procedure is stopped once convergence is reached within a prescribed tolerance.

We perform this numerical procedure without further approximations, in particular retaining the full frequency dependence of the bosonic propagators. We nevertheless choose parameters for which the auxiliary boson approximation is expected to hold, allowing for a direct comparison with the analytical prediction. Once the numerical solution for the collective fields is obtained, we extract the late--time relaxation rate from
\begin{equation}
    \Gamma = - \Im \Sigma_{\rm R}(\omega) \Big|_{\omega = 0} \ .
\end{equation}
The results are shown in Fig.~\ref{fig:Relaxation_rate_p_2} for $p = 2$ and different values of $\gamma$ (dotted lines), together with the large--$p$ analytical prediction~\eqref{eq:Late_time_relaxation_rate}. Despite being derived in the large--$p$ limit, the analytical result shows remarkable agreement with the numerical solution. The agreement is particularly accurate for $\kappa/\Delta$ large, while it remains quantitatively good also in the limit $\kappa\to0$, where the discrepancy with the large--$p$ prediction is at most $\simeq10\%$.

\section{Dissipative Lyapunov growth}
\label{sec:Lyapunov-growth}

While the late--time relaxation rate provides useful information about correlation decay, it does not by itself distinguish chaotic dynamics from other mechanisms leading to relaxation.
A more refined diagnostic of many--body chaos is provided by the late--time behavior of out-of-time-order correlators (OTOCs), whose exponential growth in chaotic systems captures the quantum analogue of the butterfly effect and defines the quantum Lyapunov exponent $\lambda$.

The behavior of the Lyapunov exponent in dissipative SYK models has recently been investigated in the presence of Markovian dissipation~\cite{Bhattacharjee_2024, Bhattacharjee2023OperatorGrowth, Garcia-Garcia:2024tbd, Liu:2024stj}. In particular, Ref.~\cite{Garcia-Garcia:2024tbd} studied the SYK model with linear fermionic jump operators and found that, while the relaxation rate $\Gamma$ generally increases with the dissipation strength $\mu$, the Lyapunov exponent changes sign at a finite critical value $\mu_{\rm c}$. This leads to a clear distinction between two regimes: for $\mu<\mu_c$, OTO perturbations grow exponentially, whereas for $\mu>\mu_c$ they decay exponentially, signaling a transition from chaotic to nonchaotic behavior. This highlights the importance of studying the Lyapunov exponent, in addition to relaxation rates, when characterizing dissipative quantum chaos.

Consistently with~\cite{Garcia-Garcia:2024tbd} (see also~\cite{PhysRevB.97.161114, PhysRevA.99.033816}), we consider as dissipative OTOC the four--fold correlation function
\begin{multline}
    \frac{1}{N^2} \sum_{i,j = 1}^N \Tr \big[ \mathsf{T}_{\mathcal C} \, \psi^j_{4_-}(t_2) \psi^i_{3_+}(0) \psi^j_{2_-}(t_1) \psi^i_{1_+}(0) \rho_{\rm ss} \big] \\
    = \mathcal G(t_1, t_2) + \frac{1}{N} \mathcal F(t_{1}, t_{2}) + \dots,
    \label{eq:four_point_function_OTOC_definition}
\end{multline}
namely a four--point function defined on a four--folded SK contour $\mathcal C$, extending the two--folded contour used for ordinary two--point functions. Here we use the contour ordering $\mathcal C = 4_-\succ3_+\succ2_-\succ1_+$, whose definition can be found in Appendix~\ref{app:Lyapunov_exponent}. In the large--$N$ limit, this four--point function can be studied perturbatively. At late times, but before the scrambling time $t \sim \log(N)$, it takes the form
\begin{equation}
    \mathcal F(t_1, t_2) = c \, e^{\lambda \frac{t_1 + t_2}{2} } f(t_1 - t_2) + \dots \ .
    \label{eq:Definition_Lyapunov}
\end{equation}
In~\eqref{eq:Definition_Lyapunov}, the constant $c$ is a real $\mathcal O(1)$ number, while $\lambda$ is the Lyapunov exponent. For quantum chaotic systems, $\lambda$ is conjectured to be positive, although positivity alone is not sufficient to establish chaos~\cite{PhysRevLett.124.140602}.

For generic many--body systems, computing the Lyapunov exponent analytically or numerically is challenging. However, for the broad class of SYK models at large-$N$, standard techniques allow one to determine $\lambda$ both analytically and numerically, which we will exploit in the following sections~\cite{Maldacena:2016hyu, Marcus:2018tsr, Kim:2019lwh}.

\subsection{The ladder kernel of Yukawa SYK}
\label{sec:Lyapunov-growth_ladder-kernel}

We study the four--point function~\eqref{eq:four_point_function_OTOC_definition} perturbatively at large--$N$ using the standard ladder diagram technique. 
Extensive details are provided in Appendix~\ref{app:Ladder_kernel}; here we briefly summarize the main steps. 

The leading contribution to equation~\eqref{eq:four_point_function_OTOC_definition} is obtained by contracting the two fields carrying index $j$ with one another, and separately the two fields carrying index $i$. Since the two index sums remain independent, this contribution is of order one and reads
\begin{multline}
    \mathcal G(t_1,t_2) = G_{4_-2_-}(t_2,t_1) G_{3_+ 1_+}(0,0) \\
    = G_>(0)G_>(t_2-t_1).
    \label{eq:disconnected_piece_4pt_function_revised}
\end{multline}
The first crossed contraction instead requires $i=j$. It therefore contains only one free index sum and is suppressed by $1/N$. With the convention in equation~\eqref{eq:four_point_function_OTOC_definition}, its contribution is
\begin{multline}
    \mathcal F_0(t_1,t_2) = G_{4_-3_+}(t_2,0)G_{2_-1_+}(t_1,0) \\
      -G_{4_-1_+}(t_2,0)G_{2_-3_+}(t_1,0) \\
    = G_>(t_1) G_>(t_2) - G_<(t_1) G_>(t_2)  \ .
    \label{eq:connected_piece_start_of_recursion_revised}
\end{multline}
This seed contribution fixes the normalization of the ladder series, but, as we show below, it does not affect the homogeneous equation determining the Lyapunov exponent.

Interactions generate additional ladder rungs, leading to a recursive structure. There are two types of contributions to the rung~\cite{Marcus:2018tsr, Kim:2019lwh}. The first one contains a propagating bosonic line connecting the two fermionic rails (together with additional fermionic propagators for $p>2$), and has the diagrammatic form
\begin{equation}
    \mathsf{K}_{b}(t_1, t_2; t_3, t_4) \; = \; \;  \text{\includegraphics[width=2.8cm, valign=c]{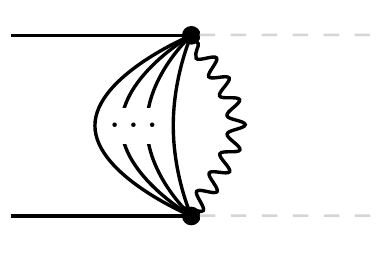}} \; \ .
\end{equation}
The subscript $b$ indicates that the rung contains a propagating bosonic mode, and we refer to the corresponding contribution as the {\it bosonic kernel}. The second type of contribution contains purely fermionic propagators in the rung, while the bosonic degrees of freedom appear through non-local retarded rails
\begin{equation}
    \mathsf{K}_{f}(t_1, t_2; t_3, t_4) \; = \; \;  \text{\includegraphics[width=2.7cm, valign=c]{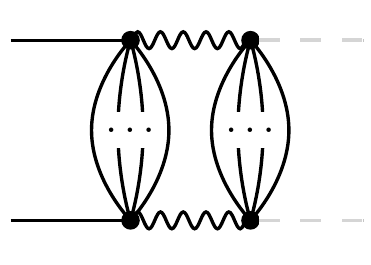}} \; \ .
\end{equation}
Once again, the subscript $f$ denotes the propagating fermions in the rung. We refer to this contribution as the {\it fermionic kernel}. All propagators entering these diagrams are understood to be dressed, consistently with the polarization diagrams~(\ref{eq:Sigma_polarization_diagram}--\ref{eq:Pi_polarization_diagram}). For simplicity, dressing corrections are not explicitly displayed in the diagrams to avoid clutter. For the derivation of these kernels from quadratic fluctuations of the effective action, we refer to Appendix~\ref{app:Ladder_kernel}. The resulting bosonic and fermionic kernels are
\begin{widetext}
\begin{align}
    \mathsf K_{b}(t_1, t_2; t_3, t_4) \, = & \, \; - \ii^{p-1} 2 \Delta \gamma J^2  (p-1) \,  G_{\rm R}(t_{13}) G_{\rm R}(t_{24}) \,  D_{>}(t_{34}) G^{p-2}_{>}(t_{34}) \ , \label{eq:Kb_main_text} \\
    \mathsf K_{f}(t_1, t_2; t_3, t_4) \, = & \, \; 4 \gamma \Delta^2 J^4 \int \de t_5 \de t_6 \, G_{\rm R}(t_{15}) G_{\rm R}(t_{26}) D_{\rm R}(t_{53}) D_{\rm R}(t_{64}) G^{p-1}_{>}(t_{34}) G^{p-1}_{>}(t_{56}) \ ,\label{eq:Kf_main_text}
\end{align}
\end{widetext}
where we have introduced $t_{ij}=t_i-t_j$. The full ladder kernel is the sum
\begin{equation}
    \mathsf{K}(t_1, t_2; t_3, t_4) = \mathsf{K}_{b}(t_1, t_2; t_3, t_4) + \mathsf{K}_{f}(t_1, t_2; t_3, t_4) \ .
\end{equation}
The leading connected four--point function therefore satisfies the Bethe--Salpeter equation
\begin{multline}
    \mathcal F = \mathcal F_0 + \mathsf K \circ \mathcal F_0 + \mathsf K \circ \mathsf K \circ \mathcal F_0 + \dots \\
    = \mathcal F_0 + \mathsf K \circ \mathcal F \ ,
    \label{eq:OTOC_BSE_definition}
\end{multline}
where the symbol $\circ$ denotes convolution over the two internal time variables,
\begin{equation}
    \mathsf K \circ \mathcal F = \int \de t_3 \de t_4 \, \mathsf K(t_1, t_2; t_3, t_4) \mathcal F(t_3, t_4) \ .
\end{equation}
At late times, we are interested in the exponentially growing contribution to the four--point function. Since the seed term $\mathcal F_0$ decays with time, we can neglect it and reduce the Bethe--Salpeter equation~\eqref{eq:OTOC_BSE_definition} to 
\begin{equation}
    \mathcal F = \mathsf K \circ \mathcal F \ .
    \label{eq:Lyapunov_comes_from_unit_eigenvalue}
\end{equation}
This homogeneous equation implies that the late-time four--point function is obtained from eigenfunctions of the ladder kernel with unit eigenvalue. It is straightforward to verify that these eigenfunctions can be parametrized as
\begin{equation}
    \mathcal F_{\lambda}(t_1,t_2) = e^{\lambda \frac{t_1 + t_2}{2}} \, f(t_1-t_2) \ .
    \label{eq:exponential_ansatz_eigenfunction}
\end{equation}
We therefore determine the Lyapunov exponent by finding the value of $\lambda$ for which the ansatz~\eqref{eq:exponential_ansatz_eigenfunction} solves the eigenvalue equation~\eqref{eq:Lyapunov_comes_from_unit_eigenvalue}.

\subsection{Enhanced scrambling at \texorpdfstring{large--$p$}{large-p}}
\label{sec:Lyapunov-growth_large-p}

We now solve the eigenvalue equation~\eqref{eq:Lyapunov_comes_from_unit_eigenvalue} using the large--$p$ saddle-point solutions obtained in Sec.~\ref{sec:Late-time_relaxation_large-p} and the ansatz~\eqref{eq:exponential_ansatz_eigenfunction}. This equation is an integral equation for the function $f(t)$. By taking one derivative with respect to each of the time variables $t_1$ and $t_2$, it can be transformed into a second-order differential equation.

As shown in Appendix~\ref{app:from_kernel_to_Poschl_Teller}, using the large--$p$ solution of Sec.~\ref{sec:Late-time_relaxation_large-p} and approximating $G_{\rm R}(t) \approx - \ii \theta(t) e^{- \Gamma t}$ as in~\cite{Garcia-Garcia:2024tbd} \footnote{We can resort to this approximation because, in the limit of $p$ large, the constant $\mathsf C$ in~\eqref{eq:late_time_relaxation_definition} is very close to one. In particular, at large--$p$ there is a factor of $p^{\frac{1}{p}} = 1+ \frac{1}{p} \log(p) + \dots$, and we neglect the subleading term in the RHS. This is consistent for $p \to \infty$, but for $p = 2$ the error is $\sim 10 \%$, thus consistent with our numerical findings.}, the resulting equation takes the form
\begin{equation}
     - \frac{\de^2}{\de t^2} f(t) + V(t) f(t) = - \bigg( \Gamma + \frac{\lambda}{2} \bigg)^2 \, f(t) \ ,
\label{eq:Lyapunov_quantum_mechanics_problem}
\end{equation}
where the effective potential is
\begin{equation}
     V(t) = - \frac{2 \mathsf{A}^{2}}{\cosh^{2}(\mathsf{A} |t| + \mathsf{B})} - \frac{2(p-1)}{p} \, \mathsf{A} \tanh(\mathsf{B})  \delta(t)  
    \label{eq:Poschl-Teller_potential_main_text}
\end{equation}
and consists of a Pöschl--Teller contribution supplemented by an attractive contact term at the origin. We emphasize that deriving~\eqref{eq:Lyapunov_quantum_mechanics_problem} requires several approximations, some of which are controlled only in the strict large--$p$ limit. Further details are provided in Appendix~\ref{app:from_kernel_to_Poschl_Teller}.

The Lyapunov exponent is determined by the lowest-energy bound state of~\eqref{eq:Lyapunov_quantum_mechanics_problem}. Dissipation affects this effective quantum-mechanical problem in two distinct ways. First, it modifies the two-point functions entering the kernel, in particular through the relaxation rate $\Gamma$ appearing on the RHS of~\eqref{eq:Poschl-Teller_potential_main_text}. Second, it directly contributes an attractive contact term to the potential, which increases the binding energy and therefore enhances the Lyapunov exponent. This term originates from the bosonic Keldysh correlator~\eqref{eq:large-p_bosonic_Keldysh_correlator}, encoding the quantum fluctuations of the dissipative bath. Hence, even though the fermionic sector is already at infinite temperature, fluctuations of the bosonic environment can actively enhance Lyapunov growth. The final behavior of $\lambda$ results from the competition between these two mechanisms.

\begin{figure*}[t]
    \centering
    \includegraphics[width=0.95\textwidth]{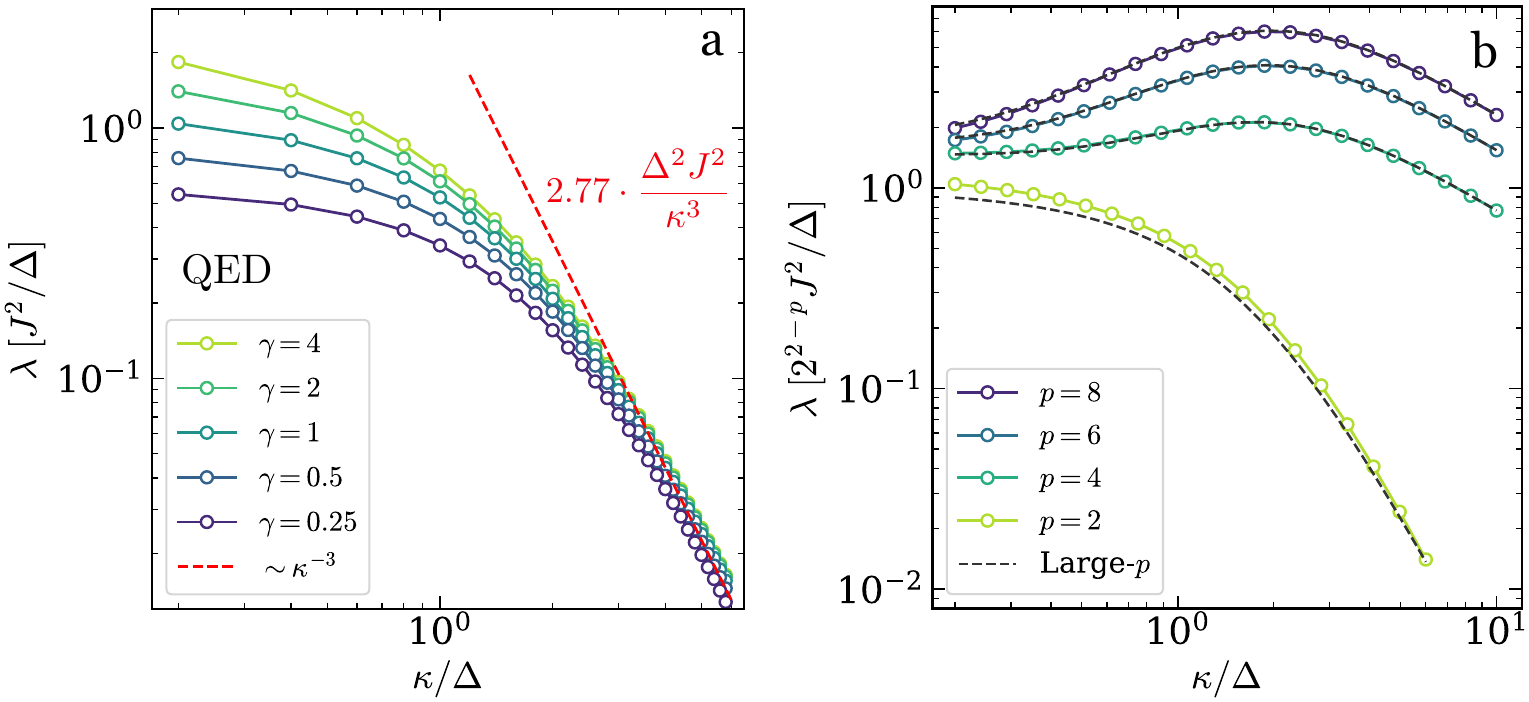}
    \caption{\textbf{a}) Dissipative Lyapunov exponent for $p = 2$, computed numerically. The case $p=2$ is the QED vertex, thus is the most relevant for platforms such as cavity--QED and circuit--QED. As predicted by the large--$p$ result, the Lyapunov exponent is always positive, it is monotonically decreasing in $\kappa$, and for large dissipation it becomes independent of $\gamma$, decaying as $\kappa^{-3}$. \textbf{b}) Dissipative Lyapunov exponent for $\gamma = 1$ and different values of $p$, compared to the large--$p$ auxiliary result. For $p = 2$, $\lambda$ is monotonically decreasing and highly suppressed at large $\kappa$. For $p>2$ and $\gamma > 2/p^2$, scrambling is enhanced until $\kappa = \kappa^*$ as in~\eqref{eq:kappa_star_definition}, and then decreases as $\sim \kappa^{-1}$.}
    \label{fig:Lyapunov_p_2}
\end{figure*}

The potential~\eqref{eq:Poschl-Teller_potential_main_text} admits a single bound state, whose wavefunction (up to normalization) is
\begin{equation}
    f(t) =  e^{- \mathsf{E} ( |t| + \mathsf{B} / \mathsf{A} )}  \Big( \mathsf{E} + \mathsf{A} \tanh \big( \mathsf{A} |t| + \mathsf{B} \big) \Big) \ ,
    \label{eq:eigenfunction_Poschl-Teller_potential}
\end{equation}
where we have defined $\mathsf E = \Gamma + \frac{\lambda}{2} \,$. The corresponding energy is determined by imposing the discontinuity condition across the $\delta$--function, namely
\begin{equation}
    f'(0^+) - f'(0^-) = - \frac{2(p-1)}{p}  \mathsf{A} \tanh(\mathsf{B})   f(0) \ .
\end{equation}
This gives the solution
\begin{equation}
    \mathsf E = \mathsf A \ \frac{\pm \sqrt{4 p^2 - (4p - 1) \tanh^2(\mathsf B)} - \tanh(\mathsf B)}{2 p} \ ,
    \label{eq:energy_bound_state}
\end{equation}
and we can only consider the positive branch, as we want a normalizable solution~\eqref{eq:eigenfunction_Poschl-Teller_potential}. Using~\eqref{eq:energy_bound_state} and the definitions~\eqref{eq:large-p_definitions_A_B}, we obtain 
\begin{multline}
    \lambda = \frac{1}{p} \, \bigg( \sqrt{ \big( 2p \big)^2 J_{\rm eff}^2 + \big( 2p - 1 \big)^2 \Gamma_{\rm Pur}^2 }  \\
    - 2 \sqrt{J_{\rm eff}^2 + \Gamma_{\rm Pur}^2 } - \Gamma_{\rm Pur} \bigg) \ .
    \label{eq:Dissipative_Lyapunov_exponent_large_p_analytics}
\end{multline}
This analytical expression is one of the central results of this work. We now examine how the predicted Lyapunov exponent behaves for various dissipation strengths.

\textit{Small dissipation} --- Expanding~\eqref{eq:Dissipative_Lyapunov_exponent_large_p_analytics} for $\kappa \ll \Delta$, we obtain 
\begin{equation}
    \lambda = \frac{(p-1)\sqrt{\gamma} \, J^2}{2^{p-3} p \, \Delta } - \frac{\gamma J^2 \kappa}{2^{p-1} \Delta^2} + \dots \ .
\end{equation}
At $\kappa=0$, this expression reduces to the high-temperature Lyapunov exponent of YSYK. For small but finite dissipation, the Lyapunov exponent is suppressed by a correction linear in $\kappa$, which is parametrically controlled by $\Gamma_{\rm Pur}/p$. This behavior is consistent with the general expectation that dissipation tends to inhibit Lyapunov growth~\cite{Chen:2017dbb, Schuster:2022bot, PhysRevLett.131.220404, PhysRevLett.130.250401, Bhattacharjee2023OperatorGrowth, Garcia-Garcia:2024tbd, Liu:2024stj}. It is interesting to notice that the formula~\eqref{eq:Dissipative_Lyapunov_exponent_large_p_analytics} reproduces a negative Lyapunov exponent for the integrable case $p = 1$, vanishing in the limit $\kappa \to 0$ \footnote{In principle, the whole ladder kernel does not make sense for $p=1$, as the four point functions identically vanishes because the system is quadratic. This implies that the Lyapunov exponent for $p = 1$ is strictly zero. However, it is interesting to notice that~\eqref{eq:Dissipative_Lyapunov_exponent_large_p_analytics} still predicts a non--scrambling behavior for $p=1$ at any $\kappa$.}.

\textit{Large dissipation} --- In the limit where $\kappa \gg \Delta$, we can simply expand~\eqref{eq:Dissipative_Lyapunov_exponent_large_p_analytics} using the fact that in this limit $\Gamma_{\rm Pur} \gg J_{\rm eff}$, leading to
\begin{equation}
    \lambda  = \frac{2(p-2)}{p} \, \Gamma_{\rm Pur} + \frac{(2 p-1)^2 + 1}{2p(2p-1)} \, \frac{J_{\rm eff}^2}{\Gamma_{\rm Pur}} + \dots \ .
    \label{eq:Large_kappa_Lyapunov_prediction}
\end{equation}
The large--dissipation expansion reveals a qualitative distinction between the cases $p>2$ and $p=2$. For the former, the large--dissipation regime is dominated by $\Gamma_{\rm Pur}$, which vanishes as $\kappa^{-1}$. In particular, we have
\begin{equation}
    \lambda_{p>2}  = \frac{(p-2) \gamma J^2}{2^{p-4} \, \kappa} + \dots \ .
    \label{eq:Large_kappa_Lyapunov_prediction_p_larger_2}
\end{equation}
In contrast, for $p=2$, the Lyapunov exponent remains positive, but it is strongly inhibited by dissipation, vanishing as
\begin{equation}
    \lambda_{p=2}  = \frac{10}{3} \cdot\frac{\Delta^2 J^2}{\kappa^3} + \dots \ .
    \label{eq:Large_kappa_Lyapunov_prediction_p_2}
\end{equation}
Besides the suppressed $\kappa^{-3}$ scaling, another interesting feature of~\eqref{eq:Large_kappa_Lyapunov_prediction_p_2} is the absence of $\gamma$ in the leading term. This cancellation occurs because $J_{\rm eff} \propto \sqrt{\gamma}$ while $\Gamma_{\rm Pur} \propto \gamma$, so the combination in~\eqref{eq:Large_kappa_Lyapunov_prediction} shows no dependence on $\gamma$ at this order.
Therefore, in this regime, the leading large--$\kappa$ behavior for $p=2$ is independent of the boson--to--fermion ratio $\gamma$.

\textit{Intermediate dissipation} --- The regime $\kappa\simeq2\Delta$ displays the most nontrivial behavior. In particular, for $p>2$ the Lyapunov exponent can become non--monotonic as a function of the dissipation strength. Indeed, let us start by fixing $\gamma = \mathcal O(1)$, independent of $p$. Then, expanding the expression~\eqref{eq:Dissipative_Lyapunov_exponent_large_p_analytics} of $\lambda$ around $\kappa = 2 \Delta$, and taking the derivative $\partial_{\kappa} \lambda$, we observe that the latter vanishes at 
\begin{align}
    \kappa^* = & \; \,  2 \Delta \Big( 1 - \frac{1}{\gamma p^2} + \dots \Big) \ , \label{eq:kappa_star_definition} \\
    \text{and} \; \; \lambda(\kappa^*) = & \; \, \frac{\gamma J^2}{2^{p-2}\Delta} \, \Big( p - 2 + \frac{1}{2 \gamma p} + \dots \Big) \ .
\end{align}
This extremum corresponds to a local maximum, which for $\gamma=\mathcal O(1)$ is also the global maximum. Therefore, in this intermediate dissipation regime, the Lyapunov exponent is enhanced over a broad range of parameters.
This holds for any $p>2$, although our analysis above hides the fact that for $p = 2$ the Lyapunov exponent is monotonically decreasing also in this intermediate regime. On the other hand, we also stress that for $p > 2$, in order to have a monotonically decreasing $\lambda$, we need to take $\gamma$ to scale parametrically with $p^2$. Doing so, we can obtain the critical $\gamma$ expanding~\eqref{eq:Large_kappa_Lyapunov_prediction} for $\kappa$ small, finding where the second derivative vanishes. We obtain
\begin{equation}
    \gamma_{\rm c} = \frac{2}{p^2} + \dots \ .
\end{equation}
Below this value, the Lyapunov exponent decreases monotonically with increasing dissipation, whereas above it an enhancement regime appears for $p>2$. Interestingly, this critical value of $\gamma$ coincides, at leading order, with the corresponding critical value~\eqref{eq:monotonicity_relaxation_rate_gamma} for the relaxation rate $\Gamma$ --- a remarkable property of our dissipative model.

\subsection{Numerical evaluation of the Lyapunov exponent} \label{sec:Lyapunov-growth_numerics}

After analyzing the analytical prediction~\eqref{eq:Dissipative_Lyapunov_exponent_large_p_analytics} in the strict auxiliary limit, we now turn to a numerical evaluation of the Lyapunov exponent. As in Sec.~\ref{sec:Late-time_relaxation_p-2}, we solve the full Schwinger--Dyson equations without imposing the auxiliary approximation, although we choose parameters for which this limit provides a controlled comparison.

The numerical strategy follows the same logic as the analytical derivation of~\eqref{eq:Dissipative_Lyapunov_exponent_large_p_analytics}. We solve the eigenvalue equation~\eqref{eq:Lyapunov_comes_from_unit_eigenvalue} by searching for an eigenfunction $\mathcal F$ of the form~\eqref{eq:exponential_ansatz_eigenfunction}. For a fixed value of $\lambda$, we compute the largest eigenvalue of the ladder kernel $\mathsf K$ and adjust $\lambda$ until this eigenvalue reaches one. The largest eigenvalue is efficiently obtained by repeated applications of $\mathsf K$. Moreover, since the kernel acts through convolutions in time, the multidimensional integrations can be efficiently evaluated in frequency space. This procedure allows for an efficient numerical extraction of the Lyapunov exponent.

We first compare the numerical results with the large--$p$ analytical prediction. We begin with the case $p=2$, shown in Fig.~\ref{fig:Lyapunov_p_2} \textbf{a}. As discussed in Sec.~\ref{sec:Lyapunov-growth_large-p}, the analytical result predicts that the Lyapunov exponent remains positive for all values of $\kappa$, decreases monotonically with increasing dissipation, and is strongly suppressed in the large--$\kappa$ regime, where it vanishes as $\kappa^{-3}$. These features are confirmed by the numerical calculation. In particular, at large $\kappa$, the numerical result is consistent with the fit
\begin{equation}
    \lambda_{p = 2, {\rm fit}} = 2.77 \cdot \frac{\Delta^2 J^2}{\kappa^3} \ ,
\end{equation}
which differs by approximately $20\%$ from the large--$p$ coefficient $10/3$. As in~\eqref{eq:Large_kappa_Lyapunov_prediction_p_larger_2}, the leading behavior is independent of $\gamma$.

We next consider different values of $p$, with numerical results shown in Fig.~\ref{fig:Lyapunov_p_2}\textbf{b}. Our goal is to test the prediction that, at fixed $\gamma$, intermediate dissipation can enhance the Lyapunov exponent for $p>2$, while the $p=2$ case exhibits a qualitatively different behavior. We compute the Lyapunov exponent for $p=2,4,6$ and $8$ at fixed $\gamma=1$. To compare different interaction orders using the same energy scale, we fix $2^{2-p} J^2 = 2.5 \times 10^{-3}$ and $\Delta=1$. The numerical results confirm the large--$p$ prediction: the agreement is excellent, and the Lyapunov exponent is enhanced for intermediate values of the dissipation strength $\kappa$.

\section{Discussion}
\label{sec:Conclusion}

In this work, we have studied the effect of bosonic dissipation on chaotic observables of the Yukawa--SYK model, described by a Markovian master equation with bosonic leakage as Lindbladian jump operators. In particular, we have investigated how dissipation modifies the late--time fermionic relaxation rate and the Lyapunov growth. We have found a rich landscape of dynamical behaviors controlled by the boson--to--fermion ratio $\gamma = R/N$ and the dissipation strength $\kappa$. Most notably, while the system remains scrambling even at arbitrarily large dissipation strengths, the Lyapunov exponent can, for suitable parameters (in particular for $p>2$), increase as $\kappa$ is enhanced. This behavior contrasts with the intuition that stronger dissipation should always suppress chaotic dynamics and highlights the nontrivial role played by the dissipative environment~\cite{Chen:2017dbb, Schuster:2022bot, PhysRevLett.131.220404, PhysRevLett.130.250401, Bhattacharjee2023OperatorGrowth, Garcia-Garcia:2024tbd, Liu:2024stj}.

The persistence, and in some regimes enhancement, of $\lambda$ can be traced back to quantum fluctuations of the dissipative bosonic sector. These fluctuations are encoded in the bosonic Keldysh propagator and are subsequently inherited by the greater and lesser correlators entering the Schwinger--Dyson equations. Thus, the bath does not simply act as a source of decoherence: its quantum fluctuations can actively contribute to the growth of perturbations. Related mechanisms for bath-induced enhancement of quantum chaos have recently been discussed in Refs.~\cite{PhysRevLett.131.160202,liu2026inducingenhancingmanybodyquantum,zhang2026enhancing}; here, we identify such an effect in a dissipative setting directly motivated by quantum-simulation platforms such as cavity QED.

At a broader level, this work addresses the question of how dissipation modifies quantum-chaotic signatures in experimentally relevant many-body systems. In noisy quantum simulators, a central challenge is to determine when environmental coupling overwhelms the intrinsic chaotic dynamics of the target system. In other words,
\begin{quote}
    {\it When does dissipation dominate over the unitary dynamics, hindering the quantum chaotic features of the target system?}
\end{quote}
In previous studies, such as Ref.~\cite{Garcia-Garcia:2024tbd}, where SYK models coupled to linear jump operators were considered, this question admits a clear answer: sufficiently strong dissipation drives the Lyapunov exponent negative. In the present setting, however, the answer is more subtle, since the Lyapunov exponent remains positive and can even be enhanced. While weak dissipation generally suppresses Lyapunov growth, there is no universal dissipation threshold beyond which scrambling is destroyed. Instead, suitable choices of jump operators can lead to a form of {\it engineered scrambling}, where the chaotic properties of the system are modified and enhanced by tuning the dissipative environment. This opens the possibility of controlling scrambling through cavity dissipation.

We conclude this discussion by presenting several future directions that will be interesting to explore in order to corroborate and complement the findings of this work.

\textit{Additional sources of dissipation} --- One of the motivations of this work was to investigate an experimentally relevant source of dissipation, namely boson leakage. While such a mechanism is ubiquitous in quantum simulators involving bosonic mediators, it is not the only dissipative process present in realistic platforms. For example, cavity QED experiments can also involve incoherent photon scattering, as {\it e.g.} in~\cite{Ferrari:2026zkn}. Incorporating such effects is an interesting direction for future work. We leave this issue for future endeavors, noting that it might be difficult to fully incorporate its effect in a path--integral formulation due to the complicated nature of the disordered jump operators. As a first step, one could investigate Lindbladians with random jump operators, such as~\cite{PhysRevB.106.075138, PhysRevResearch.4.L022068}, as simplified models of photon scattering processes.

\textit{The peculiar case of $p = 2$} --- Our results indicate that the case $p=2$ occupies a special position among the YSYK models considered here. While the system remains scrambling for any $\kappa$ (in contrast to the integrable case $p=1$), its Lyapunov exponent displays qualitatively different behavior compared to $p>2$. This distinction can be understood from the structure of the effective jump operator obtained in the adiabatic elimination of Sec.~\ref{sec:Late-time_relaxation_large-p}, which takes the form
\begin{equation}
    a_{\mu} = \frac{\ii^{- \frac{p}{2}}}{\Delta + \ii \frac{\kappa}{2}} \, \sum_{i_{1}\dots i_p}^N J_{i_1 \dots i_p}^{\mu} \, \psi^{i_1} \dots \psi^{i_p} \ .
\end{equation}
For $p=2$, the effective jump operator is quadratic in the fermions, whereas for $p>2$ it becomes a genuine many--body operator. It is therefore natural to ask whether this distinction influences operator growth. While unitary dynamics admits a relation between operator growth and the Lyapunov exponent~\cite{parker2019}, a general framework connecting these quantities in non--unitary systems is still lacking~\cite{Schuster:2022bot, Bhattacharjee_2024, Bhattacharjee2023OperatorGrowth}. Understanding the special role of $p=2$ more deeply remains an interesting open problem.

\textit{The critical value of $\gamma$} --- For two distinct observables, namely the relaxation rate and the Lyapunov exponent, we have identified qualitatively different behaviors depending on whether the boson--to--fermion ratio is above or below the value $\gamma_{\rm c}=2/p^2$. Although this scaling with $p$ can be anticipated by comparing the energy scales $J_{\rm eff}$ and $\Gamma_{\rm Pur}$ at large $p$, the agreement of the $\mathcal O(1)$ prefactor is remarkable, given that $\Gamma$ and $\lambda$ arise from different physical observables. This suggests that the value $\gamma_{\rm c}$ may admit a deeper interpretation. 

Let us finish by pointing out that in the Double Scaled SYK (DSSYK) model~\cite{polchinski2016, berkooz2019}, one defines double scaled parameters $\lambda_{\rm DSSYK} = 2 p^2/N$ and $q = e^{- \lambda}$, which organize the combinatorial expansion of observables into an exactly solvable system. For the case at hand here, we notice that the critical number of bosons $R_{\rm c}$ has a natural rewriting as $R_{\rm c} = \gamma_{\rm c} N = 2 N/p^2 = 4/ \lambda_{\rm DSSYK}$. It is therefore interesting to ask whether this value of $\gamma$ finds a natural interpretation in a double-scaled version of the YSYK model.\\

\begingroup
\renewcommand{\addcontentsline}[3]{}
\begin{acknowledgments}

We thank Ana~Asenjo~Garcia, Rahel~Lea~Baumgartner, Jean--Philippe~Brantut, Daniel~Carney, Adolfo~del~Campo, Filippo~Ferrari, Matthew~Fisher, Philipp~Hauke, Manthos~Karydas, Benjamin~Knepper, Lucas~Sá, Adrián~Sánchez-Garrido, Julian~Sonner, Jacobus~Verbaarschot and Zhenbin~Yang for insightful discussions. PP would like to thank the members of the HologrAPh consortium for discussions and publications on related topics over the years.
PP acknowledges financial support from the Swiss National Science Foundation (Postdoc.Mobility Grant No. P500PT\_230584).
SR is supported by the National Science Foundation under Award No. DMR-2409412.
This work was performed in part at the Aspen Center for Physics, which is supported by National Science Foundation grant PHY-2210452. 
\\

\textbf{AI disclosure} --- The main ideas contained in this work are original from the Authors. LLMs such as ChatGPT 5.5 and 5.6 (OpenAI) and Claude 4.8 (Anthropic) have been used to help with writing numerical codes, checking technical aspects of analytical calculations, and in writing the manuscript. The Authors take full responsibility for everything contained in this work. \\

\textbf{Data availability} --- The codes used for the numerical results contained in Figure \ref{fig:Relaxation_rate_p_2} and \ref{fig:Lyapunov_p_2} can be found on GitHub \footnote{Repository:~\url{https://github.com/PietroPelliconi/Dissipation-enhanced_scrambling_YSYK}}. 
\end{acknowledgments}
\endgroup

\bibliography{extendedrefs}

\onecolumngrid

\appendix
\addtocontents{toc}{\protect\hideappendixsubsections}

\section{Steady state} \label{app:steady_state}

In this Appendix, we determine the steady state perturbatively in the fermion--boson coupling, without assuming any hierarchy between the bosonic detuning $\Delta$ and the loss rate $\kappa$. 
As already mentioned in the main text, this is not needed for our analysis, as the time--independent Schwinger--Dyson equations already select the solution in the steady state, by construction. However, it is interesting to understand the precise form of the steady state, and its dependence on photon loss and disorder.

We want to solve for $\mathcal L(\rho_{\rm ss}) = 0$ as a perturbative expansion $\rho_{\rm ss} = \rho_0 + \rho_1 + \dots$, where each order $\rho_n = \mathcal O(J^n)$. It is convenient to introduce the Hermitian fermionic operators
\begin{equation}
    \mathcal Q_\mu \equiv \ii^{\frac{p}{2}}
    \sum_{i_1<\cdots<i_p}J^\mu_{i_1\cdots i_p}
    \psi^{i_1}\cdots\psi^{i_p}  \ ,
\end{equation}
so that
\begin{equation}
    H = \sum_{\mu=1}^R \Delta \, a_\mu^\dagger a_\mu + \sum_{\mu=1}^R \big( a_\mu + a_\mu^\dagger \big) \, \mathcal Q_\mu \ .
\end{equation}
Since we want to find the steady state perturbatively for $J \ll \{ \Delta, \kappa \}$, it is convenient to separate the Liouvillian into two pieces, namely $\mathcal L = \mathcal L_0 + \mathcal L_1$, with
\begin{equation}
    \mathcal L_0(\rho) = - \ii \sum_{\mu = 1}^R \Delta \, \big[ a_\mu^\dagger a_\mu , \rho \big]
    + \kappa \sum_{\mu=1}^R \Big(a_\mu\rho a_\mu^\dagger - \frac{1}{2} \big\{ a_\mu^\dagger a_\mu, \rho \big \} \Big) \ , \qquad \mathcal L_1(\rho) =  - \ii \sum_{\mu = 1}^R  \big[ \big( a_\mu + a_\mu^\dagger \big) \, \mathcal Q_\mu , \, \rho \, \big] \ .
\end{equation}
This is useful, as the operators $\mathcal Q_\mu$ are of order $J$, so that $\mathcal L_1=\mathcal O(J)$. Let
\begin{equation}
    \ket{0}\equiv\bigotimes_{\mu=1}^R\ket{0_\mu} \, ,
    \qquad \text{and} \qquad \ket{1_\mu}\equiv a_\mu^\dagger\ket{0} \ .
\end{equation}
At zeroth order, the steady state is
\begin{equation}
    \rho_0 = \ketbra{0}{0} \otimes  \frac{1}{2^{N/2}} \, \mathbbm 1  \ , \qquad \qquad \mathcal L_0\big(\rho_0 \big)=0 \ ,
\end{equation}
as mentioned in~\eqref{eq:Initial_density_matrix}. The first correction is obtained from $\mathcal L_0 \big( \rho_1 \big) = - \mathcal L_1 \big( \rho_{0} \big)$. Using the fact that
\begin{equation}
    \mathcal L_1 \big( \rho_0 \big) = - \ii \, \Big( \! \ketbra{1_{\mu}}{0} - \ketbra{0}{1_{\mu}} \! \Big) \otimes \frac{1}{2^{N/2}} \, \mathcal Q_{\mu} \ ,
\end{equation}
and noticing that
\begin{equation}
    \mathcal L_0 \big( \ketbra{1_\mu}{0} \big) = - \Big( \frac{\kappa}{2} + \ii \Delta \Big)  \ketbra{1_\mu}{0} \ , \qquad \qquad \mathcal L_0 \big(\ketbra{0}{1_\mu} \big) = - \Big( \frac{\kappa}{2} - \ii \Delta \Big)  \ketbra{0}{1_\mu} \ ,
\end{equation}
it is straightforward to find that the steady state, up to the first non--trivial order, is
\begin{equation}
    \rho_{\rm ss} = \ketbra{0}{0} \otimes \frac{1}{2^{N/2}}  \mathbbm 1  - \sum_{\mu=1}^R \Bigg( \frac{\ketbra{1_\mu}{0}}{\Delta - \frac{\ii \kappa}{2}} + \frac{\ketbra{0}{1_\mu}}{\Delta + \frac{\ii \kappa}{2}} \Bigg) \otimes \frac{1}{2^{N/2}} \, \mathcal Q_\mu + \dots  \ .
    \label{eq:steady_state_perturbative}
\end{equation}
This expression is Hermitian and normalized, since the correction is off diagonal in boson number and thus traceless.
Therefore, the presence of the coherent interaction between fermions and bosons produce a very weak off--diagonal component proportional to $\mathcal O_{\mu}$. Moreover, equation~\eqref{eq:steady_state_perturbative} retains the full dependence on $\Delta$ and $\kappa$ and consequently applies both when $\Delta\gg\kappa$ and when $\kappa\gg\Delta$, provided $J\ll\sqrt{\Delta^2+\kappa^2/4}$.

\section{Path integrals} \label{app:path-integrals}

In this Appendix, we show how the Yukawa-SYK model can be efficiently studied in real time by employing path integral methods, in particular the Schwinger-Keldysh formalism. The subject is broad, and we will not be able to review it here. We will mostly show schematically various results. For a more extensive discussion we refer to many review and textbook on the subject, such as~\cite{Sieberer2016Keldysh, Kamenev_2011}.

\subsection{Schwinger--Keldysh techniques}
\label{app:SK_technology}

The Schwinger--Keldysh path integral represents the trace of the time-evolved
density matrix on a closed real-time contour. Ordinary two--point functions require the usual two-branch Schwinger--Keldysh contour.
\begin{equation}
    \mathcal C=
    \mathcal C_{+}\cup\mathcal C_{-} \ , \qquad \text{with} \qquad - \succ + \ .
\end{equation}
The symbol $\succ$ reminds us that the {\it minus} branch comes after the {\it plus} branch, in the contour ordering. The case of unitary evolution is particularly simple and instructing. The evolution of a partition function without sources is
\begin{equation}
    Z=\operatorname{Tr}\!\left[U(t',t) \, \rho \, U^\dagger(t',t)\right] = \int \prod_{\sigma = \pm} \mathcal D\phi_\sigma \, \exp \Big( \ii \sum_{\sigma = \pm}\eta_{\sigma} S[\phi_\sigma] \Big) = 1 \ ,
\end{equation}
with the fields on the two branches identified at the turning point $t'$, and whose {\it orientation factors} are $\eta_{\pm} = \pm1$. In the Lindblad problem the dissipative influence functional couples the two branches, but the same closed-contour structure and branch
labels remain applicable. In the next Section we will show how the Lindbladian master equation~\eqref{eq:Lindbladian_dyanamics_generic_master_equation} can be written in the Schwinger--Keldysh formalism, and subsequently as an effective theory of bilocal two--point function. Here we mention a few generic facts that will turn out to be useful for the subsequent analysis. 

As mentioned above, the various fields $\phi$ of the system carry an index based on whether they correspond to a forward or a backward evolution, or similarly if they are inserted in the forward or backward branch. Thus, one can consider generic correlation functions with a generic number of fields insertions in both the forward or backward contour. However, it is quite common to reduce such higher--order correlation functions into (convolutions of) two--point functions, {\it e.g.} exploiting perturbation theory, carrying two indices such as $F_{\pm \pm}(t,t')$. Another basis commonly used is the $cq$--basis, which is defined as
\begin{equation}
    \phi_{c} = \frac{\phi_+ + \phi_-}{\sqrt2} \ , \qquad \qquad \phi_{q} = \frac{\phi_+ - \phi_-}{\sqrt2} \ .
\end{equation}
The $cq$--basis is arguably the most convenient basis when dealing with causal correlators. Indeed, it is not hard to show that
\begin{equation}
    \sum_{\sigma, \sigma'} \eta_{\sigma} \eta_{\sigma'} \, x_{\sigma} F_{\sigma \sigma'} x_{\sigma'} = x_+ F_{++} x_+ - x_+ F_{+-} x_- - x_- F_{-+} x_+ + x_- F_{--} x_- = x_{q} F_{\rm R} x_{c} + x_{c} F_{\rm A} x_{q} + x_{q} F_{\rm K} x_{q} \ ,
\end{equation}
defining the {\it retarded}, {\it advanced} and {\it Keldysh} components of a two--point function. They are related to the $\pm$ branches as
\begin{align}
    F_{++} \, = & \; \frac{1}{2} \big( F_{\rm K} + F_{\rm A} + F_{\rm R} \big) \ ,  \\
    F_{+-} \, = & \; \frac{1}{2} \big( F_{\rm K} + F_{\rm A} - F_{\rm R} \big) \ ,  \\
    F_{-+} \, = & \; \frac{1}{2} \big( F_{\rm K} - F_{\rm A} + F_{\rm R} \big) \ ,  \\
    F_{--} \, = & \; \frac{1}{2} \big( F_{\rm K} - F_{\rm A} - F_{\rm R} \big) \ .
\end{align}
From the definitions above, it is also simple to find that
\begin{equation}
    F_{\rm A} =  F_{++} - F_{-+}  \ ,  \qquad  F_{\rm R} =  F_{++} - F_{+-}  \ , \qquad F_{\rm K} =  F_{++} + F_{--} = F_{+-} + F_{-+} \ ,
\end{equation}
and to realize that there is the redundancy
\begin{equation}
    F_{++} + F_{--} - F_{+-} - F_{-+} = 0 \ ,
\end{equation}
which holds both in equilibrium and out-of-equilibrium. Finally, the recurrent combination
\begin{equation}
    \sum_{\sigma, \sigma'} \eta_{\sigma} \eta_{\sigma'} \, F_{\sigma \sigma'} H_{\sigma' \sigma} = F_{++} H_{++} - F_{+-} H_{-+} - F_{-+} H_{+-} + F_{--} H_{--} = F_{\rm A} H_{\rm A} + F_{\rm R} H_{\rm R}
    \label{eq:product_of_bilocals}
\end{equation}
has a very simple rewriting in terms of advanced and retarded functions, and will reveal useful later.

\subsection{Yukawa--SYK effective action}
\label{app:YSYK_effective_action}

We want to use the Schwinger-Keldysh path integral formalism to study the dynamics of our dissipative system. We start from the path integral
\begin{equation}
    Z = \int \prod_{\mu=1}^R \mathcal D \big[ x_\pm^\mu , \pi_\pm^\mu \big] \prod_{i=1}^N  \mathcal D \big[ \psi_\pm^i \big] \, e^{\ii S [x, \pi, \psi ]}
\end{equation}
 where the action is
\begin{multline}
    \ii S [a, \psi ] = \ii \int \de t \, \Bigg( \sum_{\sigma = \pm } \eta_{\sigma} \sum_{\mu = 1}^R \Big( \pi^\mu_\sigma(t) \dot x^\mu_\sigma(t) - \frac{1}{2} \, \pi^\mu_\sigma(t) \pi^\mu_\sigma(t) - \frac{\Delta^2}{2} \, x_\sigma^{\mu}(t) x_\sigma^{\mu}(t) \Big) + \frac{1}{2} \sum_{\sigma = \pm } \eta_{\sigma} \sum_{i=1}^N \psi^i_\sigma(t) \, \ii \, \dot \psi^i_\sigma(t) \Bigg) \\
    + \frac{\kappa}{2\Delta} \int \de t \sum_{\mu = 1}^R \bigg( - \frac{\Delta^2}{2} \big( x_+^\mu(t) - x_-^\mu(t) \big)^2 - \frac{1}{2} \big( \pi_+^\mu(t) - \pi_-^\mu(t) \big)^2  -\ii \Delta x_+^\mu(t) \pi_-^\mu(t) + \ii \Delta x_-^\mu(t) \pi_+^\mu(t) \bigg) \\  
    - \ii^{\frac{p}{2}+1} \int \de t \sqrt{2 \Delta} \sum_{\sigma = \pm } \eta_{\sigma} \sum_{\mu=1}^R \sum_{i_{1}\dots i_p}^N J_{i_1 \dots i_p}^{\mu}  x_\sigma^{\mu}(t) \,  \psi^{i_1}_\sigma(t) \dots \psi^{i_p}_\sigma(t)  \ .
\end{multline}
This is the path--integral rewriting of the Lindblad master equation. 

We now take the disorder average of the partition function to obtain an effective bilocal field theory. Using a Gaussian probability density function with zero mean and variance set by~\eqref{eq:variance_of_disordered_couplings}, the disorder average is a simple Gaussian integral. We then obtain
\begin{equation}
    Z_{\rm eff} \equiv \mathbb E \big[ Z \big] = \int \prod_{\mu=1}^R \mathcal D \big[ x_\pm^\mu , \pi_\pm^\mu \big] \prod_{i=1}^N  \mathcal D \big[ \psi_\pm^i \big] \, e^{\ii S_{\rm eff} [x, \pi, \psi ]} \ ,
\end{equation}
with the effective action being
\begin{multline}
    \ii S_{\rm eff}[x,\pi, \psi] = \ii \int \de t \, \Bigg( \sum_{\sigma = \pm } \eta_{\sigma} \sum_{\mu = 1}^R \Big( \pi^\mu_\sigma(t) \dot x^\mu_\sigma(t) - \frac{1}{2} \, \pi^\mu_\sigma(t) \pi^\mu_\sigma(t) - \frac{\Delta^2}{2} \, x_\sigma^\mu(t) x_\sigma^{\mu}(t) \Big) - \frac{1}{2} \sum_{\sigma = \pm } \eta_{\sigma} \sum_{i=1}^N \psi^i_\sigma(t) \, \ii \, \dot \psi^i_\sigma(t) \Bigg) \\
    + \frac{\kappa}{2\Delta} \int \de t \sum_{\mu = 1}^R \Big( - \frac{\Delta^2}{2} \big( x_+^\mu(t) - x_-^\mu(t) \big)^2 - \frac{1}{2} \big( \pi_+^\mu(t) - \pi_-^\mu(t) \big)^2  -\ii \Delta x_+^\mu(t) \pi_-^\mu(t) + \ii \Delta x_-^\mu(t) \pi_+^\mu(t) \Big) \\
    + (-1)^{\frac{p}{2} + 1} \int \de t \, \de t' \frac{(p-1)! J^2 \Delta}{N^p} \sum_{\sigma,\sigma'} \eta_{\sigma} \eta_{\sigma'} \sum_{\mu=1}^R \sum_{i_{1}\dots i_p}^N x_\sigma^{\mu}(t) \, x_{\sigma'}^{\mu}(t') \, \psi^{i_1}_\sigma(t) \dots \psi^{i_p}_\sigma(t) \psi^{i_1}_{\sigma'}(t') \dots \psi^{i_p}_{\sigma'}(t')   \ .
    \label{eq:disorder_averaged_effective_action_x_coordinates}
\end{multline}
This is the effective disorder averaged effective action.

\subsection{Bilocal collective fields}
\label{app:SK_collective_fields}

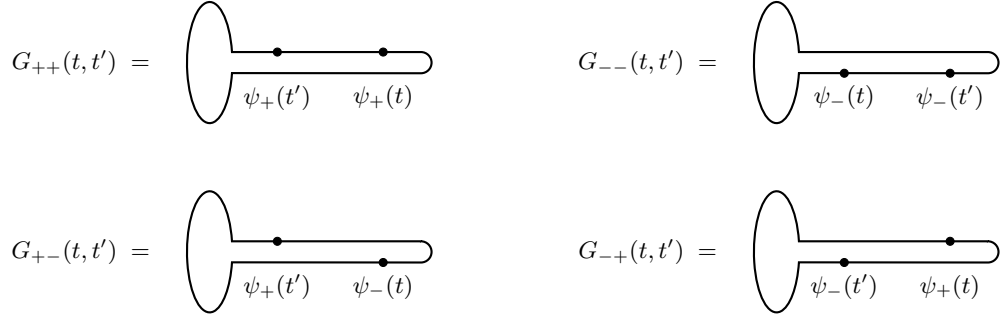
\begin{figure}[t]
\centering
\begin{tikzpicture}
    \draw[black, thick] (2.5,0) -- (0,0) arc [start angle=10, end angle=350, x radius=0.3, y radius=0.8] -- (2.5,-0.28) -- (2.5,-0.28) arc [start angle=-90, end angle=90, x radius=0.14, y radius=0.14] -- cycle;
    \filldraw (0.6,0) circle (1.5pt);
    \filldraw (2,0) circle (1.5pt);

    \draw[] (0.6,-0.6) node {$\psi_+(t')$};
    \draw[] (2,-0.6) node {$\psi_+(t)$};
    \draw[] (-2,-0.14) node {$G_{++} (t, t') \; = $};

\begin{scope}[xshift=7.5cm]
    \draw[black, thick] (2.5,0) -- (0,0) arc [start angle=10, end angle=350, x radius=0.3, y radius=0.8] -- (2.5,-0.28) -- (2.5,-0.28) arc [start angle=-90, end angle=90, x radius=0.14, y radius=0.14] -- cycle;
    \filldraw (0.6,-0.28) circle (1.5pt);
    \filldraw (2,-0.28) circle (1.5pt);

    \draw[] (0.6,-0.6) node {$\psi_-(t)$};
    \draw[] (2,-0.6) node {$\psi_-(t')$};
    \draw[] (-2,-0.14) node {$G_{--} (t, t') \; = $};
\end{scope}

\begin{scope}[yshift=-2.5cm]
    \draw[black, thick] (2.5,0) -- (0,0) arc [start angle=10, end angle=350, x radius=0.3, y radius=0.8] -- (2.5,-0.28) -- (2.5,-0.28) arc [start angle=-90, end angle=90, x radius=0.14, y radius=0.14] -- cycle;
    \filldraw (0.6,0) circle (1.5pt);
    \filldraw (2,-0.28) circle (1.5pt);

    \draw[] (0.6,-0.6) node {$\psi_+(t')$};
    \draw[] (2,-0.6) node {$\psi_-(t)$};
    \draw[] (-2,-0.14) node {$G_{+-} (t, t') \; = $};
\end{scope}

\begin{scope}[xshift=7.5cm, yshift=-2.5cm]
    \draw[black, thick] (2.5,0) -- (0,0) arc [start angle=10, end angle=350, x radius=0.3, y radius=0.8] -- (2.5,-0.28) -- (2.5,-0.28) arc [start angle=-90, end angle=90, x radius=0.14, y radius=0.14] -- cycle;
    \filldraw (0.6,-0.28) circle (1.5pt);
    \filldraw (2,0) circle (1.5pt);

    \draw[] (0.6,-0.6) node {$\psi_-(t')$};
    \draw[] (2,-0.6) node {$\psi_+(t)$};
    \draw[] (-2,-0.14) node {$G_{-+} (t, t') \; = $};
\end{scope}
\end{tikzpicture}
\caption{Schwinger--Keldysh real--time correlators. As in the main text, we define $G_{++}(t, t') \equiv \mathsf T G(t, t')$ the {\it time ordered} two--point function,  $G_{--}(t, t') \equiv \tilde{\mathsf T} G(t, t')$ the {\it anti--time ordered} one, while the {\it greater} and the {\it lesser} are $G_{-+}(t, t') \equiv G_{>}(t, t')$ and $G_{+-}(t, t') \equiv G_{<}(t, t')$, respectively. The closing circles represent the trace.}
\label{fig:SK_real_time_correlators}
\end{figure}

We are now in the position to introduce the bilocal collective field variables. For the Majorana fermions we use the definitions
\begin{align}
    G_{++}(t, t') \, = & \; G_{\sfT}(t, t') = -\frac{\ii}{N} \sum_{i = 1}^N \sfT \psi^i_+(t) \psi^i_+(t') \ , & & G_{--}(t, t') \, = \; G_{\sfaT}(t, t') = -\frac{\ii}{N} \sum_{i = 1}^N \sfaT \psi^i_-(t) \psi^i_-(t') \ , \\
    G_{-+}(t, t') \, = & \; G_{>}(t, t') = -\frac{\ii}{N} \sum_{i = 1}^N \psi^i_-(t) \psi^i_+(t') \ , & & G_{+-}(t, t') \, = \; G_{<}(t, t') = - \frac{\ii}{N} \sum_{i = 1}^N \psi^i_+(t) \psi^i_-(t') \ .
\end{align}
In the expressions above, the operator $\sfT$ is the time ordering (fermionic) operator, while $\tilde \sfT$ is the anti--time ordering one. In particular, for fermions they are defined as
\begin{align}
    \sfT\psi(t_1) \psi(t_2) \, = & \; \theta(t_1-t_2) \, \psi(t_1) \psi(t_2) - \theta(t_2-t_1) \, \psi(t_2) \psi(t_1) \ , \\
    \tilde \sfT\psi(t_1) \psi(t_2) \, = & \; \theta(t_1-t_2) \, \psi(t_2) \psi(t_1) - \theta(t_2-t_1) \, \psi(t_1) \psi(t_2) \ .
\end{align}
We then also define the contour--ordered bosonic two--point functions as
\begin{align}
    D_{++}(t, t') \, = & \; D_{\sfT}(t, t') = -\frac{\ii}{R} \sum_{\mu = 1}^R \sfT x_+^\mu(t) x_+^\mu(t') \ , && D_{--}(t, t') \, = \; D_{\sfaT}(t, t') = -\frac{\ii}{R} \sum_{\mu = 1}^R \sfaT x_-^\mu(t) x_-^\mu(t') \ , \\
    D_{-+}(t, t') \, = & \; D_{>}(t, t') = -\frac{\ii}{R} \sum_{\mu = 1}^R x_-^\mu(t) x_+^\mu(t') \ , && D_{+-}(t, t') \, = \; D_{<}(t, t') = - \frac{\ii}{R} \sum_{\mu = 1}^R x_+^\mu(t') x_-^\mu(t) \ .
\end{align}
In this case the bosonic time ordering operator $\sfT$ and the anti--time ordering operator $\tilde \sfT$ are defined as
\begin{align}
    \sfT x(t_1) x(t_2) \, = & \; \theta(t_1-t_2) \, x(t_1) x(t_2) + \theta(t_2-t_1) \, x(t_2) x(t_1) \ , \\
    \tilde \sfT x(t_1) x(t_2) \, = & \; \theta(t_1-t_2) \, x(t_2) x(t_1) + \theta(t_2-t_1) \, x(t_1) x(t_2) \ .
\end{align}
These are our definitions of the collective field variables of the effective action. To rewrite the path integral in terms of these variables we can employ the identity \footnote{To ease the notation in~\eqref{eq:introduction_of_self_energy_fermions} and~\eqref{eq:introduction_of_self_energy_bosons} we have neglected the time ordering and anti--time ordering operators. When $(\sigma, \sigma') = (+,+)$ or $(-,-)$ they are present, of course.}
\begin{multline}
    1 = \int \mathcal D G_{\sigma \sigma'}(t, t') \, \delta \Big( G_{\sigma \sigma'}(t,t') + \frac{\ii}{N} \sum_{i = 1}^N \psi^i_\sigma(t) \psi^i_{\sigma'}(t') \Big) \\
    = \int \mathcal D \Sigma_{\sigma' \sigma}(t',t) \, \mathcal D G_{\sigma \sigma'}(t,t') \exp \bigg(  \frac{N}{2} \, \eta_{\sigma} \eta_{\sigma'} \int \de t \, \de t' \, \Sigma_{\sigma' \sigma}(t',t) \Big( G_{\sigma \sigma'}(t,t') + \frac{\ii}{N} \sum_{i = 1}^N \psi^i_\sigma(t) \psi^i_{\sigma'}(t') \Big) \bigg) \ ,
    \label{eq:introduction_of_self_energy_fermions}
\end{multline}
up to a multiplication constant. Similarly for the bosonic variables we have
\begin{multline}
    1 = \int \mathcal D D_{\sigma \sigma'}(t, t') \, \delta \Big( D_{\sigma \sigma'}(t,t') + \frac{\ii}{R} \sum_{\mu = 1}^R x^\mu_\sigma(t) x^\mu_{\sigma'}(t') \Big) \\
    = \int \mathcal D \Pi_{\sigma' \sigma}(t',t) \, \mathcal D D_{\sigma \sigma'}(t,t') \exp \bigg( - \frac{R}{2} \, \eta_{\sigma} \eta_{\sigma'} \int \de t \, \de t' \, \Pi_{\sigma' \sigma}(t',t) \Big( D_{\sigma \sigma'}(t,t') + \frac{\ii}{R} \sum_{\mu = 1}^R x_\mu^\sigma(t) x_\mu^{\sigma'}(t') \Big) \bigg) \ ,
    \label{eq:introduction_of_self_energy_bosons}
\end{multline}
up to a multiplication constant as before. We can now introduce these identities into the disorder averaged partition function, and use the $\delta$-function identifications to rewrite the bilocal interaction term as
\begin{multline}
    (-1)^{\frac{p(p-1)}{2}+1} \frac{(p-1)! J^2 \Delta}{N^p} \sum_{\sigma,\sigma'} \eta_{\sigma} \eta_{\sigma'} \sum_{\mu=1}^R \sum_{i_{1}\dots i_p}^N x^\sigma_{\mu}(t) \, x^{\sigma'}_{\mu}(t') \, \psi^{i_1}_\sigma(t) \dots \psi^{i_p}_\sigma(t) \psi^{i_1}_{\sigma'}(t') \dots \psi^{i_p}_{\sigma'}(t') \\
    \to \quad (-\ii)^{p+1} \frac{R J^2 \Delta}{p} \sum_{\sigma,\sigma'} \eta_{\sigma} \eta_{\sigma'} D_{\sigma \sigma'}(t,t') G^p_{\sigma \sigma'}(t,t') \ .
\end{multline}
We are now left with Gaussian integrals to perform, on bosonic and fermionic variables. Let us first concentrate on the bosonic ones. We first do the bosonic integral over momenta, and then we perform the one over positions. Thus we first tackle
\begin{multline}
    \int \prod_{\mu = 1}^R \mathcal D \big[ \pi_{\pm}^{\mu} \big] \, \exp \Bigg( \ii \int \de t \, \Bigg( \sum_{\mu = 1}^R \Big( \pi^\mu_+ \dot x^\mu_+ - \pi^\mu_- \dot x^\mu_- - \frac{1}{2} \, \pi^\mu_+ \pi^\mu_+ + \frac{1}{2} \, \pi^\mu_- \pi^\mu_- - \frac{\Delta^2}{2} \, x^\mu_+ x^\mu_+ + \frac{\Delta^2}{2} \, x^\mu_- x^\mu_- \Big)  \Bigg) \\
    + \int \de t \, \frac{\kappa}{2\Delta} \sum_{\mu = 1}^R \Big( - \frac{\Delta^2}{2} \big( x_+^\mu - x_-^\mu \big)^2 - \frac{1}{2} \big( \pi_+^\mu - \pi_-^\mu \big)^2 + \ii \Delta \pi_+^\mu x_-^\mu - \ii \Delta \pi_-^\mu  x_+^\mu \Big) \Bigg) \ .
\end{multline}
To simplify our life we use the $cq$--basis
\begin{align}
    x_c^\mu(t) = \frac{x_+^{\mu}(t) + x_-^{\mu}(t)}{\sqrt{2}} \ , & &   x_q^\mu(t) = \frac{x_+^{\mu}(t) - x_-^{\mu}(t)}{\sqrt{2}} \ , \\
    \pi_c^\mu(t) = \frac{\pi_+^{\mu}(t) + \pi_-^{\mu}(t)}{\sqrt{2}} \ , & &   \pi_q^\mu(t) = \frac{\pi_+^{\mu}(t) - \pi_-^{\mu}(t)}{\sqrt{2}} \ .
\end{align}
Let us also neglect the time variable in the various fields, in order to ease the notation. We obtain
\begin{equation}
    \int \prod_{\mu = 1}^R \mathcal D \big[ \pi_{cq}^{\mu} \big] \, \exp \Bigg( \ii \int \de t \, \bigg( \sum_{\mu = 1}^R \Big(\pi^\mu_c \dot x^\mu_q + \pi^\mu_q \dot x^\mu_c - \pi^\mu_c \pi^\mu_q - \Delta^2 x^\mu_c x^\mu_q  + \ii \frac{\kappa}{2\Delta} \big( ({\pi_q^\mu})^2 + \Delta^2 ({x_q^\mu})^2 \big)  + \frac{\kappa}{2} \big( \pi_q^\mu x_c^\mu  -  \pi_c^\mu  x_q^\mu \big) \Big) \bigg) \Bigg) \ .
    \label{eq:bosonic_Gaussian_path_integral_cq_variables}
\end{equation}
In the above integral, the term proportional to $\pi_c^{\mu}(-\omega)$ is linear, which is then a representation of a $\delta$-function setting
\begin{equation}
    \pi_q^{\mu} - \dot x_q^{\mu} + \frac{\kappa}{2} \, x_q^\mu = 0 \ .
\end{equation}
The remaining path integral over $\pi_q^{\mu}$ can be easily solved substitute the relation above~\eqref{eq:bosonic_Gaussian_path_integral_cq_variables}, finding
\begin{equation}
    \exp \Bigg( \ii \int \de t \, \sum_{\mu = 1}^R \bigg( \Big( \dot x_q^{\mu} - \frac{\kappa}{2} \, x_q^\mu \Big) \dot x^\mu_c - \Delta^2 x^\mu_c x^\mu_q + \ii \frac{\kappa}{2\Delta} \Big( \Big( \dot x_q^{\mu} - \frac{\kappa}{2} \, x_q^\mu \Big)^2 + \Delta^2 {x_q^\mu}^2 \Big)  + \frac{\kappa}{2} \Big( \dot x_q^{\mu} - \frac{\kappa}{2} \, x_q^\mu \Big) x_c^\mu   \bigg)  \Bigg)  \ .
    \label{eq:bosonic_Gaussian_path_integral_p_final}
\end{equation}
Let us rewrite this part of the action in frequency space. We obtain
\begin{equation}
    \exp \Bigg( \frac{\ii}{2} \int \de \omega \, \sum_{\mu = 1}^R \begin{pmatrix}
        x^{\mu}_c(-\omega) \\
        x^{\mu}_q(-\omega)
    \end{pmatrix}^{\! \! \mathsf T } \begin{pmatrix}
        0 & \big[ D^0_{\rm A} \big]^{-1}(\omega) \\
        \big[ D^0_{\rm R} \big]^{-1}(\omega) & \big[ D^0_{\rm K} \big]^{-1}(\omega)
    \end{pmatrix} \begin{pmatrix}
        x^{\mu}_c(\omega) \\
        x^{\mu}_q(\omega)
    \end{pmatrix}  \Bigg)  \ ,
    \label{eq:bosonic_Gaussian_path_integral_p_final_frequency}
\end{equation}
where
\begin{align}
     \big[ D^0_{\rm R} \big]^{-1}(\omega) \, = & \; \Big( \omega + \frac{\ii \kappa}{2} \Big)^2 - \Delta^2 \ , \\
    \big[ D^0_{\rm A} \big]^{-1}(\omega) \, = & \; \Big( \omega - \frac{\ii \kappa}{2} \Big)^2 - \Delta^2 \ , \\
    \big[ D^0_{\rm K} \big]^{-1}(\omega) \, = & \; \frac{\ii \kappa}{\Delta}\Big( \omega^2 + \frac{\kappa^2}{4} + \Delta^2 \Big)  \ .
\end{align}
are the inverse dissipative free bosonic propagators. From this expression it is also simple to obtain the propagators from the inverse matrix, namely
\begin{equation}
    \begin{pmatrix}
        0 & \big[ D^0_{\rm A} \big]^{-1}(\omega) \\
        \big[ D^0_{\rm R} \big]^{-1}(\omega) & \big[ D^0_{\rm K} \big]^{-1}(\omega)
    \end{pmatrix}^{-1} = \; \begin{pmatrix}
        D^0_{\rm K} (\omega) &  D^0_{\rm R} (\omega) \\
        D^0_{\rm A} (\omega) &  0
    \end{pmatrix}  \ ,
\end{equation}
with
\begin{align}
    D^0_{\rm R} (\omega) \, = & \; \frac{1}{\big( \omega + \frac{\ii \kappa}{2} \big)^2 - \Delta^2} \ , \\
    D^0_{\rm A} (\omega) \, = & \; \frac{1}{\big( \omega - \frac{\ii \kappa}{2} \big)^2 - \Delta^2} \ , \\
    D^0_{\rm K} (\omega) \, = & \; - \frac{\ii \kappa}{2\Delta}\bigg( \frac{1}{\big( \omega - \Delta \big)^2 + \frac{\kappa^2}{4}} + \frac{1}{\big( \omega + \Delta \big)^2 + \frac{\kappa^2}{4}} \bigg)  \ .
\end{align}
We now need to add the effect of the interaction with the fermions, namely the bosonic self energy. We can rewrite
\begin{equation}
    - \frac{\ii}{2} \sum_{\sigma, \sigma'} \eta_{\sigma} \eta_{\sigma'} \int \de t \, \de t' \, \Pi_{\sigma' \sigma}(t',t) \sum_{\mu = 1}^R x_\mu^\sigma(t) x_\mu^{\sigma'}(t')  = - \frac{\ii}{2} \, \int \de \omega \, \begin{pmatrix}
        x^{\mu}_c(-\omega) \\
        x^{\mu}_q(-\omega)
    \end{pmatrix}^{\! \! \mathsf T } \begin{pmatrix}
        0 & \Pi_{\rm A} (\omega) \\
        \Pi_{\rm R}(\omega) & \Pi_{\rm K}(\omega)
    \end{pmatrix} \begin{pmatrix}
        x^{\mu}_c(\omega) \\
        x^{\mu}_q(\omega)
    \end{pmatrix}
\end{equation}
All in all, the Gaussian integral over bosonic variables yields 
\begin{multline}
    \int \prod_{\mu = 1}^R \mathcal D \big[ x_{cq}^{\mu} \big] \, \exp \Bigg( \frac{\ii}{2} \int \de \omega \,  \sum_{\mu = 1}^R  \begin{pmatrix}
        x^{\mu}_c(-\omega) \\
        x^{\mu}_q(-\omega)
    \end{pmatrix}^{\! \! \mathsf T } \begin{pmatrix}
        0 & \big[ D^0_{\rm A} \big]^{-1}(\omega) - \Pi_{\rm A}(\omega) \\
        \big[ D^0_{\rm R} \big]^{-1}(\omega) - \Pi_{\rm R}(\omega) & \big[ D^0_{\rm K} \big]^{-1}(\omega) - \Pi_{\rm K}(\omega)
    \end{pmatrix} \begin{pmatrix}
        x^{\mu}_c(\omega) \\
        x^{\mu}_q(\omega)
    \end{pmatrix} \Bigg) \\
    = \exp \Bigg( - \frac{R}{2} \Tr \log \begin{pmatrix}
        0 & \big[ D^0_{\rm A} \big]^{-1}(\omega) - \Pi_{\rm A}(\omega) \\
        \big[ D^0_{\rm R} \big]^{-1}(\omega) - \Pi_{\rm R}(\omega) & \big[ D^0_{\rm K} \big]^{-1}(\omega) - \Pi_{\rm K}(\omega)
    \end{pmatrix}  \Bigg) \equiv \exp \Bigg( - \frac{R}{2} \Tr \log \Big( \mathbf D_0^{-1}(\omega) - \mathbf \Pi(\omega) \Big) \Bigg) \ ,
    \label{eq:bosonic_Gaussian_path_integral_cq_variables_final_path_I}
\end{multline}
up to a normalization constant which we neglect. This is the {\it kinetic part} of the bosonic effective action, relating the bosonic propagators to the interactions via the self--energies. 

Let us now perform a similar calculation for the fermions, which is inevitably simpler due to the absence of direct dissipative terms, but presents some well-known due to the fermionic statistic. In particular we can write again
\begin{multline}
    \int \prod_{i = 1}^N \mathcal D \big[ \psi^i \big] \exp \Bigg( \frac{\ii}{2} \int \de \omega \sum_{i = 1}^N \begin{pmatrix} 
        \psi_{c}^i(-\omega) \\
        \psi_{q}^i(-\omega)
    \end{pmatrix}^{\! \! \mathsf T} \begin{pmatrix}
        0 && \big[G^0_{\rm A}]^{-1}(\omega) - \Sigma_{\rm A}(\omega) \\
        \big[G^0_{\rm R}]^{-1}(\omega) - \Sigma_{\rm R}(\omega) && \big[G^0_{\rm K}]^{-1}(\omega) - \Sigma_{\rm K}(\omega)
    \end{pmatrix} \begin{pmatrix} 
        \psi_{c}^i(-\omega) \\
        \psi_{q}^i(-\omega)
    \end{pmatrix} \Bigg) \\
    = \bigg( \operatorname{Pf} \Big[ \, \ii \, \Big( \mathbf G_0^{-1}(\omega) - \mathbf \Sigma(\omega) \Big) \Big] \bigg)^{N} \ ,
\end{multline}
where $\operatorname{Pf} \big[ \, \cdot \, \big]$ is the Pfaffian of a matrix. We can now use the fact that 
\begin{equation}
    \bigg( \operatorname{Pf} \Big[ \, \ii \, \Big( \mathbf G_0^{-1}(\omega) - \mathbf \Sigma(\omega) \Big) \Big] \bigg)^{N} = \bigg( \det \Big[ \, \ii \, \Big( \mathbf G_0^{-1}(\omega) - \mathbf \Sigma(\omega) \Big) \Big] \bigg)^\frac{N}{2} = \exp \bigg( \frac{N}{2} \Tr \log \Big[ \, \ii \, \Big( \mathbf G_0^{-1}(\omega) - \mathbf \Sigma(\omega) \Big) \Big] \bigg) \ .
\end{equation}
All in all, bringing one factor of $N$ to the LHS and using the definition $\gamma = R/N$, the bilocal effective action is
\begin{multline}
    \frac{\ii}{N} \, S_{\rm eff} \big[ G \Sigma D \Pi \big] = \frac{1}{2} \Tr \log \Big( \mathbf G_0^{-1}(\omega) - \mathbf \Sigma(\omega) \Big) - \frac{\gamma}{2} \Tr \log \Big( \mathbf D_0^{-1}(\omega) - \mathbf \Pi(\omega) \Big) \\
    + \sum_{\sigma, \sigma'} \frac{1}{2} \, \eta_{\sigma} \eta_{\sigma'} \int \de t \, \de t' \, \Sigma_{\sigma' \sigma}(t',t)  G_{\sigma \sigma'}(t,t') - \frac{\gamma}{2} \, \eta_{\sigma} \eta_{\sigma'} \int \de t \, \de t' \, \Pi_{\sigma' \sigma}(t',t)  D_{\sigma \sigma'}(t,t') \\
    +(-\ii)^{p+1} \frac{\gamma J^2 \Delta}{p} \int \de t \, \de t' \sum_{\sigma,\sigma'} \eta_{\sigma} \eta_{\sigma'} D_{\sigma \sigma'}(t,t') G^p_{\sigma \sigma'}(t,t') \ ,
    \label{eq:Bilocal_effective_action_final}
\end{multline}
As usual in SYK, the bilocal effective action has an overall extensive factor of $N$ which allows us to study the theory at large--$N$ in the saddlepoint approximation~\cite{Maldacena:2016hyu}. 

After quite some work, we finally arrived at the dissipative bilocal effective action~\eqref{eq:Bilocal_effective_action_final}, from which all chaotic observables of interest to us can be derived. In Sec.~\ref{app:SD_equations} we will derive the saddlepoint equations of the collective fields. In Sec.~\ref{app:Lyapunov_exponent}
 we will study the effective theory of quadratic fluctuations around the saddlepoints, and derive an effective theory for four--point functions. This will be instrumental to study OTOCs and Lyapunov growth.

\subsection{Schwinger--Dyson equations} \label{app:SD_equations}

In this section, we derive the saddlepoint equations for the effective action~\eqref{eq:Bilocal_effective_action_final}, which are also called Schwinger--Dyson equations. The first relations that we derive are the ones connecting the two--point functions with their respective self energies. Using the fact that for every matrix $\mathcal K$
\begin{equation}
    \delta \Tr \log \big( \mathcal K \big) = \Tr \big( \mathcal K^{-1} \delta \mathcal K \big) \ ,
\end{equation}
and using~\eqref{eq:product_of_bilocals}, we obtain
\begin{align}
    G_{\rm R}(\omega) \, = & \; \; \frac{1}{\omega + \ii \varepsilon - \Sigma_{\rm R}(\omega)} \ , \\
    G_{\rm A}(\omega) \, = & \; \; \frac{1}{\omega - \ii \varepsilon - \Sigma_{\rm A}(\omega)} \ , \\
    G_{\rm K}(\omega) \, = & \; G_{\rm R} (\omega) \Sigma_{\rm K}(\omega) G_{\rm A}(\omega) \ , \textcolor{white}{\Big)}
\end{align}
for the fermions, and likewise 
\begin{align}
    D_{\rm R}(\omega) \, = & \; \frac{1}{\big( \omega + \frac{\ii \kappa}{2} \big)^2 - \Delta^2 - \Pi_{\rm R}(\omega)} \ , \\ 
    D_{\rm A}(\omega) \, = & \; \frac{1}{\big( \omega - \frac{\ii \kappa}{2} \big)^2 - \Delta^2 - \Pi_{\rm A}(\omega)} \ , \\
    D_{\rm K}(\omega) \, = & \; D_{\rm R} (\omega) \Big( \Pi_{\rm K}(\omega) - \big[ D_{\rm K}^0 \big]^{-1} \Big) D_{\rm A}(\omega) \ , 
\end{align}
for the bosons. Then, we can vary the two--point functions to obtain
\begin{align}
    \Sigma_{>} (t) \, = & \; -(- \ii)^{p+1} \, 2 \Delta \gamma J^2 \, D_{<}(-t) G_{<}^{p-1}(-t) = \ii^{p-1} \, 2 \Delta \gamma J^2 \, D_{>}(t) G_{>}^{p-1}(t)  \ , \\
    \Sigma_{<} (t) \, = & \; -(- \ii)^{p+1} \, 2 \Delta \gamma J^2 \, D_{>}(-t) G_{>}^{p-1}(-t) = \ii^{p-1} \, 2 \Delta \gamma J^2 \, D_{<}(t) G_{<}^{p-1}(t) \ ,
\end{align}
and 
\begin{align}
    \Pi_{>} (t) \, = & \; (- \ii)^{p+1} \, \frac{2 \Delta J^2 }{p} \, G_{<}^{p}(-t) = - \ii^{p+1}  \, \frac{2 \Delta J^2 }{p} \, G_{>}^{p}(t)  \ , \\
    \Pi_{<} (t) \, = & \; (- \ii)^{p+1} \, \frac{2 \Delta J^2}{p}  \, G_{>}^{p}(-t) = - \ii^{p+1}  \, \frac{2 \Delta J^2 }{p} \, G_{<}^{p}(t) \ .
\end{align}
In the SD equations above, in the RHS we have also used the fact that, for stationary states (where all quantities depend only on time differences), one has $G_{>}(t) = - G_{<}(-t)$ for Majorana two--point functions, and $D_{>}(t) = D_{<}(-t)$ for bosonic two--point functions.

As we show in the main text, this systems of equations is closed, meaning that they can be used to find (in the saddlepoint approximation) their solution. We have numerically solved them via $\alpha$--iteration, and at large--$p$.

\section{The Lyapunov exponent} \label{app:Lyapunov_exponent}

In this Appendix, we study four--point functions of Majorana operators, in particular OTOCs and the Lyapunov exponent. Ordinary two--point functions require the usual two-branch Schwinger--Keldysh contour.  The OTOC, however, is represented on the four--fold contour
\begin{equation}
    \mathcal C_{\rm OTOC} = \mathcal C_{1_+} \cup \mathcal C_{2_-} \cup \mathcal C_{3_+} \cup \mathcal C_{4_-} \ , \qquad \qquad 4_-\succ3_+\succ2_-\succ1_+ \ ,
\end{equation}
whose orientation factors are $\eta_{1_+}=\eta_{3_+}=+1$ and $\eta_{2_-}=\eta_{4_-}=-1$.
The YSYK interaction functional extends to the augmented contour by allowing its contour indices to run over all four legs. 

Our strategy to obtain the leading connected part of the OTOC involves studying the quadratic fluctuations of the bilocal effective action~\eqref{eq:Bilocal_effective_action_final}. In particular, we show that such quadratic fluctuations reproduce the $1/N$ connected piece in~\eqref{eq:four_point_function_OTOC_definition}, and also systematically give the ladder diagrams mentioned in Section~\ref{sec:Lyapunov-growth}. To find this quadratic effective action, we write the saddle and its fluctuations as
\begin{equation}
    G_{\sigma \sigma'}(t,t')=\bar G_{\sigma \sigma'}(t-t')+\delta G_{\sigma \sigma'}(t,t'),
    \qquad
    D_{\sigma \sigma'}(t,t')=\bar D_{\sigma \sigma'}(t-t')+\delta D_{\sigma \sigma'}(t,t'),
    \label{eq:bilocal_fluctuations_definition}
\end{equation}
where $\bar G_{\sigma \sigma'}(t)$ and $\bar D_{\sigma \sigma'}(t)$ are the saddlepoint solutions, and analogously for $\Sigma$ and $\Pi$. Here of course $ \{ \sigma, \sigma' \}\in\mathcal C_{\rm OTOC}$. Although the stationary saddle depends only on a time difference, a generic bilocal fluctuation depends on both times separately.

The overall factor of $N$ in the collective action immediately fixes the
large--$N$ counting.  If $\delta\Phi$ collectively denotes $\delta G$ and
$\delta D$, the quadratic functional has the schematic form
\begin{equation}
    \ii S_{\rm eff} \big[ G \Sigma D \Pi \big] = \ii S_{\rm eff} \big[ \bar G \bar \Sigma \bar D \bar \Pi \big] + \frac{N}{2}\, \delta \Phi \circ \mathcal H \circ \delta \Phi + \dots \ , \qquad \big \langle \delta \Phi \, \delta \Phi \big \rangle_c =\frac{1}{N}\,\mathcal H^{-1}.
\end{equation}
In particular,
\begin{equation}
    \frac{1}{N^2}\sum_{i,j}
    \big\langle
       \psi^j_{4_-}(t_2)\psi^i_{3_+}(0)
       \psi^j_{2_-}(t_1)\psi^i_{1_+}(0)
    \big\rangle_c
    =
    \big\langle
       \delta G_{4_-2_-}(t_2,t_1)
       \delta G_{3_+1_+}(0,0)
    \big\rangle_c
    = \mathcal O \big( 1 / N \big).
\end{equation}
Thus the $1/N$ suppression of the connected four--point function is simply the Gaussian suppression of collective fluctuations.

\subsection{Quadratic fluctuations of the collective action} \label{app:quadratic_fluctuations_collective_action}

We now find the quadratic effective action and the corresponding ladder kernel. We start again from the bilocal effective action~\eqref{eq:Bilocal_effective_action_main_text}, which we rewrite here for convenience,
\begin{multline*}
    \frac{\ii}{N} \, S_{\rm eff} \big[ G \Sigma D \Pi \big] = \frac{1}{2} \Tr \log \Big( \mathbf G_0^{-1}(\omega) - \mathbf \Sigma(\omega) \Big) - \frac{\gamma}{2} \Tr \log \Big( \mathbf D_0^{-1}(\omega) - \mathbf \Pi(\omega) \Big) \\
    + \sum_{\sigma, \sigma'} \frac{1}{2} \, \eta_{\sigma} \eta_{\sigma'} \int \de t \, \de t' \, \Sigma_{\sigma' \sigma}(t',t)  G_{\sigma \sigma'}(t,t') - \frac{\gamma}{2} \, \eta_{\sigma} \eta_{\sigma'} \int \de t \, \de t' \, \Pi_{\sigma' \sigma}(t',t)  D_{\sigma \sigma'}(t,t') \\
    + (-\ii)^{p+1} \frac{\gamma J^2 \Delta}{p} \int \de t \, \de t' \sum_{\sigma,\sigma'} \eta_{\sigma} \eta_{\sigma'} D_{\sigma \sigma'}(t,t') G^p_{\sigma \sigma'}(t,t') \ ,
\end{multline*}
We then expand such action in the fluctuations~\eqref{eq:bilocal_fluctuations_definition} up to quadratic order, neglecting the linear piece which vanishes on-shell. We have
\begin{multline}
    \frac{1}{2} \Tr \log \Big( \mathbf G_0^{-1} - \bar{\mathbf \Sigma} - \delta \mathbf \Sigma  \Big) = \frac{1}{2} \Tr \log \Big(\bar{\mathbf G}^{-1} - \delta \mathbf \Sigma  \Big) = \frac{1}{2} \Tr \log \Big(\bar{\mathbf G}^{-1} \Big) + \frac{1}{2} \Tr \log \Big(1 - \bar{\mathbf G} \circ \delta \mathbf \Sigma  \Big) \\
    = \frac{1}{2} \Tr \log \Big(\bar{\mathbf G}^{-1} \Big) - \frac{1}{2} \Tr \Big(\bar{\mathbf G} \circ \delta \mathbf \Sigma  \Big) - \frac{1}{4} \Tr \Big(\bar{\mathbf G} \circ \delta \mathbf \Sigma \circ \bar{\mathbf G} \circ \delta \mathbf \Sigma  \Big) + \dots \ .
\end{multline}
Similarly, we have 
\begin{equation}
    - \frac{\gamma}{2} \Tr \log \Big( \mathbf D^{-1} - \delta \mathbf \Pi \Big) = - \frac{\gamma}{2} \Tr \log \Big(\bar{\mathbf D}^{-1} \Big) + \frac{\gamma}{2} \Tr \Big(\bar{\mathbf D} \circ \delta \mathbf \Pi  \Big) + \frac{\gamma}{4} \Tr \Big(\bar{\mathbf D} \circ \delta \mathbf \Pi \circ \bar{\mathbf D} \circ \delta \mathbf \Pi  \Big) + \dots \ .
\end{equation}
Then for the $G\Sigma$ piece
\begin{multline}
    \qquad \sum_{\sigma, \sigma'} \frac{1}{2} \, \eta_{\sigma} \eta_{\sigma'} \int \de t \, \de t' \, \Sigma_{\sigma' \sigma}(t',t)  G_{\sigma \sigma'}(t,t') \supset  \sum_{\sigma, \sigma'} \frac{1}{2} \, \eta_{\sigma} \eta_{\sigma'} \int \de t \, \de t' \, \delta \Sigma_{\sigma' \sigma}(t',t)  \delta G_{\sigma \sigma'}(t,t') \\
    = \frac{1}{2} \Tr \Big( \delta \mathbf \Sigma \circ \delta \mathbf G \Big) \ , \qquad 
\end{multline}
while for the $D\Pi$ one
\begin{multline}
    \qquad \sum_{\sigma, \sigma'} - \frac{\gamma}{2} \, \eta_{\sigma} \eta_{\sigma'} \int \de t \, \de t' \, \Pi_{\sigma' \sigma}(t',t)  D_{\sigma \sigma'}(t,t') \supset  \sum_{\sigma, \sigma'} - \frac{\gamma}{2} \, \eta_{\sigma} \eta_{\sigma'} \int \de t \, \de t' \, \delta \Pi_{\sigma' \sigma}(t',t)  \delta D_{\sigma \sigma'}(t,t') \\
    = - \frac{\gamma}{2} \Tr \Big( \delta \mathbf \Pi \circ \delta \mathbf D \Big) \ . \qquad 
\end{multline}
Finally, the interaction term gives
\begin{multline}
    (-\ii)^{p+1} \frac{\gamma J^2 \Delta}{p} \int \de t \, \de t' \sum_{\sigma,\sigma'} \eta_{\sigma} \eta_{\sigma'} D_{\sigma \sigma'}(t,t') G^p_{\sigma \sigma'}(t,t') \\
    \supset (-\ii)^{p+1} \gamma J^2 \Delta \int \de t \, \de t' \sum_{\sigma,\sigma'} \eta_{\sigma} \eta_{\sigma'} G^{p-1}_{\sigma \sigma'}(t,t') \delta D_{\sigma \sigma'}(t,t') \delta G_{\sigma \sigma'}(t,t') \; + \\
    (-\ii)^{p+1} \frac{(p-1)\gamma J^2 \Delta}{2} \int \de t \, \de t' \sum_{\sigma,\sigma'} \eta_{\sigma} \eta_{\sigma'} D_{\sigma \sigma'}(t,t') G^{p-2}_{\sigma \sigma'}(t,t') \delta G_{\sigma \sigma'}(t,t') \delta G_{\sigma \sigma'}(t,t') \ .
\end{multline}
We now want to integrate out the self energies. Since we expanded only to second order, the integral to be performed is Gaussian, and up to an unimportant additive constant (to the action), the result of the integral can be evaluated on--shell. For the fermionic self energy, varying the above action with respect to $\delta \mathbf \Sigma$ gives the condition
\begin{equation}
    \delta \mathbf G = \bar{\mathbf G} \circ \delta \mathbf \Sigma \circ \bar{\mathbf G} \ , \qquad \qquad  \text{or equivalently} \qquad \qquad \delta \mathbf \Sigma = \bar{\mathbf G}^{-1} \circ \delta \mathbf G \circ \bar{\mathbf G}^{-1} \ .
\end{equation}
Likewise, varying the quadratic action in the bosonic self energy gives the condition
\begin{equation}
    \delta \mathbf \Pi = \bar{\mathbf D}^{-1} \circ \delta \mathbf D \circ \bar{\mathbf D}^{-1} \ .
\end{equation}
Introducing these two conditions back into the quadratic action, we obtain all in all (neglecting the constant contribution given by the saddlepoints)
\begin{multline}
    \frac{\ii}{N} \, S_{\rm eff} \big[ \delta G , \delta D \big] = \frac{1}{4} \Tr \Big(\bar{\mathbf G}^{-1} \circ \delta \mathbf G \circ \bar{\mathbf G}^{-1} \circ \delta \mathbf G  \Big) - \frac{\gamma}{4} \Tr \Big(\bar{\mathbf D}^{-1} \circ \delta \mathbf D \circ \bar{\mathbf D}^{-1} \circ \delta \mathbf D \Big) \\
    + (-\ii)^{p+1} \gamma J^2 \Delta \int \de t \, \de t' \sum_{\sigma,\sigma'} \eta_{\sigma} \eta_{\sigma'} G^{p-1}_{\sigma \sigma'}(t,t') \delta D_{\sigma \sigma'}(t,t') \delta G_{\sigma \sigma'}(t,t') \; + \\
    (-\ii)^{p+1} \frac{(p-1)\gamma J^2 \Delta}{2} \int \de t \, \de t' \sum_{\sigma,\sigma'} \eta_{\sigma} \eta_{\sigma'} D_{\sigma \sigma'}(t,t') G^{p-2}_{\sigma \sigma'}(t,t') \delta G_{\sigma \sigma'}(t,t') \delta G_{\sigma \sigma'}(t,t') \ ,
    \label{eq:Quadratic_Fluctuations_Final}
\end{multline}
In the next Section we use this effective action to find the ladder kernels for the OTOCs.

\subsection{Ladder kernel} \label{app:Ladder_kernel}

We now study the correlation functions of~\eqref{eq:Quadratic_Fluctuations_Final}, {\it i.e.} the connected OTOC. In principle, one could attempt writing the above quadratic fluctuation functional in matrix form, and compute the inverse matrix to find the propagator. One entry of such matrix would directly give our target correlator $\big \langle \delta G_{4_- 2_-}(t,t') \, \delta G_{3_+ 1_+}(0,0) \big \rangle$, even though it is not particularly convenient to follow this route.

A simpler way to obtain the result needed is to perturb the effective theory with a source, and then compute the linear response of the fields with respect to the source. The derivative of the field with respect to the source is the desired propagator, in a Gaussian theory. The outcome of this procedure gives a kernel, the {\it ladder kernel}, whose solution is the leading connected OTOC. To briefly summarize this simple statement, let us consider the simple integral
\begin{equation}
    Z[j] = \int \prod_{a} \de x_a \, e^{- \frac{1}{2} \, x_a A_{ab} x_b  + x_a j_a} \ .
\end{equation}
Then, the equation of motion for the field is 
\begin{equation}
    A_{ab} x_b = j_a \qquad \to \qquad x_a = \big( A^{-1} \big)_{ab} \, j_b \ .
\end{equation}
It is then clear that 
\begin{equation}
    \big \langle x_a x_b \big \rangle = \Big \langle \frac{\de x_a}{\de j_b}  \Big \rangle \Big|_{j=0} = \big( A^{-1} \big)_{ab} \ .
\end{equation}
We will use this simple fact to find the an equation expressing the inverse propagator of~\eqref{eq:Quadratic_Fluctuations_Final}. Let us assume that we couple $\delta G_{\sigma \sigma'} (t, t')$ to a source $J_{\sigma \sigma'}(t,t')$.
The equations of motion then encode how the system responds to the action of a source. Taking a derivative of~\eqref{eq:Quadratic_Fluctuations_Final} with respect to the fermionic fluctuation we obtain
\begin{equation}
    - \mathbf G^{-1} \circ \delta \mathbf G \circ \mathbf G^{-1} + \ii^{p-1} 2 \gamma J^2 \Delta \, \Big( \mathbf G^{p-1} \odot \delta \mathbf D +  (p-1) \,  \mathbf D \odot \mathbf G^{p-2} \odot \delta \mathbf G \Big)  + \mathbf J = 0  \ .
\end{equation}
Above the $\odot$ symbol signifies entry--wise multiplication, so that $\mathbf A \odot \mathbf B \odot \mathbf C \dots  = \sum_{ab} A_{ab} B_{ab} C_{ab} \dots$.We can do the same for bosons, for which we do not introduce a source. We then have
\begin{equation}
    - \mathbf D^{-1} \circ \delta \mathbf D \circ \mathbf D^{-1} + \ii^{p-1} 2 J^2 \Delta \, \mathbf G^{p-1} \odot \delta \mathbf G    = 0 \ .
\end{equation}
Solving for the field variations we have
\begin{align}
    \delta\mathbf G
    ={}&\bar{\mathbf G}\circ\mathbf J\circ\bar{\mathbf G}
    +\ii^{p-1}2\gamma J^2\Delta\,
    \bar{\mathbf G}\circ
    \Big(
        \bar{\mathbf G}^{p-1}\odot\delta\mathbf D
        +(p-1)\bar{\mathbf D}\odot
        \bar{\mathbf G}^{p-2}\odot\delta\mathbf G
    \Big)\circ\bar{\mathbf G},
    \label{eq:dG_EOM_with_source}\\
    \delta\mathbf D
    ={}&\ii^{p-1}2J^2\Delta\,
    \bar{\mathbf D}\circ
    \Big[\bar{\mathbf G}^{p-1}\odot\delta\mathbf G\Big]
    \circ\bar{\mathbf D}.
    \label{eq:dD_EOM_with_source}
\end{align}
From now on, to ease the notation, we will omit the bar on top of the on--shell solutions of the Schwinger--Dyson equations. As advertised above, we can use the results (\ref{eq:dG_EOM_with_source}-\ref{eq:dD_EOM_with_source}) to find the propagators. In particular, we want
\begin{equation}
    \mathcal F_{4_- 2_-}(t,t') = \frac{\de \delta G_{4_- 2_-}(t,t')}{\de J_{1_+ 3_+}(0,0)} \ .
    \label{eq:F42_in_terms_of_linear_response}
\end{equation}
We have decided to keep the notation of~\eqref{eq:Quadratic_Fluctuations_Final} in terms of $t$ and $t'$, instead of $t_1$ and $t_2$, to avoid a confusing with the OTOC branches. There are essentially two pieces that contribute to~\eqref{eq:F42_in_terms_of_linear_response}, as seen in~\eqref{eq:OTOC_BSE_definition}. The first one is clearly the {\it seed} contribution coming from the disconnected propagators when $i = j$, which is obtained differentiating the first term on the RHS of~\eqref{eq:dG_EOM_with_source} with respect to the source, 
\begin{equation}
    \mathcal F_{4_- 2_- }^{0}  = \frac{\de \delta G_{4_- 2_-}(t,t')}{\de J_{3_+ 1_+}(0,0)} \bigg|_{J=0} = G_{4_- 3_+}(t, 0) G_{2_- 1_+}(t', 0) - G_{4_- 1_+}(t, 0) G_{2_- 3_+}(t',0) \ .
\end{equation}
This is indeed the $1/N$ contribution that we report in~\eqref{eq:connected_piece_start_of_recursion_revised}. Overall, we find $\mathcal F_{4_- 2_-}(t,t')$ defining first
\begin{equation}
    \mathcal F_{4_- 2_-}(t,t') = \frac{\de \delta G_{4_- 2_-}(t,t')}{\de J_{3_+ 1_+}(0,0)} \ , \qquad \qquad \mathcal B_{4_- 2_-}(t,t') = \frac{\de \delta D_{4_- 2_-}(t,t')}{\de J_{3_+ 1_+}(0,0)} \ .
\end{equation}
We then have, differentiating (\ref{eq:dG_EOM_with_source}-\ref{eq:dD_EOM_with_source}) with respect to $J_{1_+ 3_+}(0,0)$
\begin{multline}
    \mathcal F_{4_- 2_-}(t, t') = \mathcal F_{4_- 2_- }^{0}(t,t')
    + \ii^{p-1} 2 \gamma J^2 \Delta \sum_{c,d} \eta_c \eta_d \int \de t_c \, \de t_d \,  G_{4_- c}(t, t_c) G_{d 2_-}(t_d, t') \, \cdot \\ 
    \Big( G^{p-1}_{cd}(t_c, t_d) \mathcal B_{cd}(t_c, t_d) +  (p-1) \,  D_{cd}(t_c, t_d) G^{p-2}_{cd}(t_c, t_d) \mathcal F_{cd}(t_c, t_d) \Big)  \ ,
    \label{eq:F_kernel_definition}
\end{multline}
for the fermionic four--point function, while
\begin{equation}
    \mathcal B_{4_- 2_-}(t, t') = \ii^{p-1} 2 J^2 \Delta \sum_{c,d} \eta_c \eta_d \int \de t_c \, \de t_d \,  D_{4_- c}(t, t_c) D_{d 2_-}(t_d, t') \, G^{p-1}_{cd}(t_c, t_d) \mathcal F_{cd}(t_c, t_d)   \ .
    \label{eq:B_kernel_definition}
\end{equation}
We can also rewrite these equations in a more compact manner, using the matrix notation~\cite{Marcus:2018tsr}
\begin{equation}
    \begin{pmatrix}
        \mathcal F \\
        \mathcal B
    \end{pmatrix} = \begin{pmatrix}
        \mathcal F^0 \\
        0
    \end{pmatrix}
    + \begin{pmatrix}
        \mathsf K_{GG} & \mathsf K_{GD} \\
        \mathsf K_{DG} & 0
    \end{pmatrix} \circ \begin{pmatrix}
        \mathcal F \\
        \mathcal B
    \end{pmatrix} \ ,
    \label{eq:kernel_equation_matrix_definition}
\end{equation}
where the definitions of the kernels follows directly from~\eqref{eq:F_kernel_definition} and~\eqref{eq:F_kernel_definition}.

The only missing ingredient to obtain the ladder kernels to compute the Lyapunov exponent is to specify the OTOC contour.
To obtain the OTOC contour, we have $c\in \big\{ 3_+, 4_- \big\}$, while $d \in \big\{ 1_+, 2_- \big\}$. This is because the former pairs with $4_-$ and the latter pairs with $2_-$ This choice leads to the fact that, for every choice of $c$ and $d$ we have
\begin{equation}
    G_{cd}(t_c, t_d) = G_{>}(t_c-t_d) \ , \qquad \qquad D_{cd}(t_c, t_d) = D_{>}(t_c-t_d) \ ,
\end{equation}
as in the OTOC contour $c \succ d$. This is for the internal rungs of the ladder. Then, for the rails we have
\begin{equation}
    \sum_{c} \eta_c \, G_{4_-c}(t, t_c) = G_{4_-3_+}(t, t_c) - G_{4_-4_-}(t, t_c) = G_{>}(t - t_c) - G_{\tilde{\mathsf T}}(t - t_c) = G_{\rm R}(t - t_c) \ ,
\end{equation}
while for the other branch\begin{equation}
    \sum_{d} \eta_d \, G_{d2_-}(t_d, t') = G_{1_+2_-}(t_d, t') - G_{2_-2_-}(t_d, t') = G_{<}(t_d - t') - G_{\tilde{\mathsf T}}(t_d - t') = G_{\rm A}(t_d - t') = - G_{\rm R}(t' - t_d) \ .
\end{equation}
A similar statement holds for bosonic propagators, with the only difference that $D_{\rm A}(t_d - t_b) = D_{\rm R}(t_b - t_d)$. Therefore we have all the necessary ingredients to properly define the OTOC ladder kernels, as
\begin{align}
    \mathsf K_{GG}(t, t' ; t_c, t_d) \, = & \; - \ii^{p-1} 2 \gamma \Delta J^2  (p-1) \,  G_{\rm R}(t-t_c) G_{\rm R}(t' - t_d) \,  D_{>}(t_c - t_d) G^{p-2}_{>}(t_c - t_d)  \ , \\
    \mathsf K_{GD}(t, t' ; t_c, t_d) \, = & \; - \ii^{p-1} 2 \gamma \Delta J^2  \,  G_{\rm R}(t-t_c) G_{\rm R}(t' - t_d) \,  G^{p-1}_{>}(t_c - t_d)  \ , \\
    \mathsf K_{DG}(t, t' ; t_c, t_d) \, = & \; \ii^{p-1} 2 \Delta J^2 \,  D_{\rm R}(t-t_c) D_{\rm R}(t' - t_d) \,  G^{p-1}_{>}(t_c - t_d)  \ .
\end{align}
We can use these definitions to find the bosonic and fermionic kernels. Indeed, from~\eqref{eq:kernel_equation_matrix_definition} we have
\begin{equation}
    \mathcal F = \mathcal F_0 + \mathsf K_{GG} \circ \mathcal F + \mathsf K_{GD} \circ \mathcal B \ , \qquad \text{and} \qquad \mathcal B = \mathsf K_{DG} \circ \mathcal F \ . 
\end{equation}
Then clearly, as we are interested in the fermionic four--point function, we have
\begin{equation}
    \mathcal F = \mathcal F_0 + \mathsf K_{GG} \circ \mathcal F + \mathsf K_{GD} \circ \mathsf K_{DG} \circ \mathcal F \equiv \mathcal F_0 + \mathsf K_{b} \circ \mathcal F + \mathsf K_{f} \circ \mathcal F \equiv \mathcal F_0 +  \mathsf K \circ \mathcal F \ .
\end{equation}
Thus, this gives
\begin{align}
    \mathsf K_{b}(t, t'; t_c, t_d) \, = & \; - \ii^{p-1} 2 \gamma \Delta J^2 (p-1) \,  G_{\rm R}(t - t_{c}) G_{\rm R}(t' - t_{d}) \,  D_{>}(t_{cd}) G^{p-2}_{>}(t_{cd}) \ , \label{eq:Kb_appendix} \\
    \mathsf K_{f}(t, t'; t_c, t_d) \, = & \, \; 4 \gamma \Delta^2 J^4 \int \de t_m \de t_n \, G_{\rm R}(t-t_{m}) G_{\rm R}(t' - t_{n}) D_{\rm R}(t_{mc}) D_{\rm R}(t_{nd}) G^{p-1}_{>}(t_{cd}) G^{p-1}_{>}(t_{mn}) \label{eq:Kf_appendix} \ .
\end{align}
which are exactly the kernels given in~\eqref{eq:Kb_main_text} and~\eqref{eq:Kf_main_text} in the main text, with a slightly more complicated notation. This conclude our technical derivation of the ladder kernels for the OTOC four--point function.

\subsection{Lyapunov exponent at \texorpdfstring{large--$p$}{large-p} and the Pöschl–Teller potential} \label{app:from_kernel_to_Poschl_Teller}

We are now tasked to find a solution for the functional equation
\begin{equation}
    \mathcal F(t_1, t_2) = \int \de t_3 \, \de t_4 \,  \mathsf K(t_1, t_2; t_3, t_4) \, \mathcal F(t_3, t_4) \ ,
    \label{eq:kernel_eigenvalue_equation_appendix}
\end{equation}
thus to find the eigenvector of the operator $\mathsf K(t_1, t_2; t_3, t_4)$ with unit eigenvalue. We consider the family of functions
\begin{equation}
    \mathcal F_{\lambda}(t_1, t_2) = e^{\lambda \frac{t_1 + t_2}{2}} \, f(t_1-t_2) \ .
    \label{eq:mathcal_F_ansatz}
\end{equation}
We then want to find the value of $\lambda$ for which~\eqref{eq:kernel_eigenvalue_equation_appendix} is solved. We use the standard strategy (see {\it e.g.}~\cite{Maldacena:2016hyu}) to transform such integral equation in a second order ODE differentiating with respect to $t_1$ and $t_2$. We also approximate 
\begin{equation}
    G_{\rm R}(t) \approx - \ii \theta(t) \, e^{- \Gamma t} \ ,
    \label{eq:retarded_function_Garcia_approximation}
\end{equation}
which holds at large-$p$~\cite{Garcia-Garcia:2024tbd}. For the sake of clarity, let us do it step by step.

We begin with the bosonic kernel. Using~\eqref{eq:Kb_appendix} and the approximation~\eqref{eq:retarded_function_Garcia_approximation}, we have
\begin{equation}
    \mathcal F_{\lambda}(t_1,t_2) = \ii^{p-1} 2 \gamma \Delta J^2 (p-1) \int_{- \infty}^{t_1} \de t_3 \int_{- \infty}^{t_2}  \de t_4 \, e^{- \Gamma  (t_1 - t_3)} \, e^{- \Gamma  (t_2 - t_4)} \,  D_{>}(t_{34}) G^{p-2}_{>}(t_{34}) \mathcal F_{\lambda}(t_3,t_4) \ .
\end{equation}
Then, taking two derivatives in $t_1$ and $t_2$, we can rewrite this integral equation as
\begin{equation}
    \big( \partial_{t_1} + \Gamma \big) \big( \partial_{t_2} + \Gamma \big) \mathcal F_{\lambda}(t_1,t_2) = \ii^{p-1} 2 \gamma \Delta J^2 (p-1) \,  D_{>}(t_{12}) G^{p-2}_{>}(t_{12}) \mathcal F_{\lambda}(t_1,t_2) \ .
    \label{eq:middle_equation_derivation_poschl_teller_potential_boson}
\end{equation}
Using now the ansatz~\eqref{eq:mathcal_F_ansatz}, the LHS becomes
\begin{equation}
    \big( \partial_{t_1} + \Gamma \big) \big( \partial_{t_2} + \Gamma \big) \mathcal F_{\lambda}(t_1,t_2) = - e^{\lambda \frac{t_1 + t_2}{2} } \, f''(t_1-t_2) + \Big( \Gamma + \frac{\lambda}{2}  \Big)^2 e^{\lambda \frac{t_1 + t_2}{2} } \, f(t_1-t_2) \ .
\end{equation}
On the other hand, using the large--$p$ solution found in Sec.~\ref{sec:Late-time_relaxation_large-p}, for the RHS of~\eqref{eq:middle_equation_derivation_poschl_teller_potential_boson} we obtain
\begin{equation}
    \ii^{p-1} 2 \gamma \Delta J^2 (p-1) \,  D_{>}(t_{12}) G^{p-2}_{>}(t_{12}) \mathcal F_{\lambda}(t_1,t_2) \approx \bigg( \frac{2 (p-1)}{p} \, \mathsf{A} \tanh(\mathsf B) \delta(t_{12}) + \frac{\mathsf{A}^2}{\cosh^2 (\mathsf A |t_{12}| + \mathsf B)} \bigg) \mathcal F_{\lambda}(t_1,t_2) \ .
\end{equation}
To obtain the above equation, we have to perform some approximations which are mutually valid only in the strict large--$p$ limit. For instance, we have kept the prefactor $(p-1)/p$ as is in the term with the $\delta$-function, while we have approximated $(p-1)/p \approx 1$ in the term proportional to $\cosh^2 (\mathsf A |t_{12}| + \mathsf B)$. Let us justify these choices employing the fact that these equations are valid in the limit of $p$ large, so interpreting such expressions at finite--$p$ is inevitably an approximation. The ones performed above are not all mutually consistent for finite--$p$, but they give neat results, which can then be compared to numerics to assess their validity. The comparison is quite favorable, as shown in Figures~\ref{fig:Relaxation_rate_p_2} and~\ref{fig:Lyapunov_p_2}, and we thus refrain from complicating the analysis. Nonetheless, a solution keeping all terms in $1/p$ can be found, either perturbatively or numerically.

Let us now switch to the fermionic kernel. We start from the fact that in the auxiliary limit, the retarded bosonic two--point functions are proportional to a $\delta$--function, namely
\begin{equation}
    D_{\rm R}(t) = - \frac{1}{\Delta^2 + \frac{\kappa^2}{4}} \, \delta(t)  \ .
\end{equation}
This implies that the fermionic kernel becomes 
\begin{equation}
    \mathsf K_{f}(t_1, t_2; t_3, t_4) = \frac{4 \gamma \Delta^2 J^4}{\big( \Delta^2 + \frac{\kappa^2}{4} \big)^2} \, G_{\rm R}(t_{13}) G_{\rm R}(t_{24}) G^{2p-2}_{>}(t_{34}) \ .
\end{equation}
Employing the same strategy that led to~\eqref{eq:middle_equation_derivation_poschl_teller_potential_boson}, we obtain
\begin{equation}
    \big( \partial_{t_1} + \Gamma \big) \big( \partial_{t_2} + \Gamma \big) \mathcal F_{\lambda}(t_1,t_2) = - \frac{4 \gamma \Delta^2 J^4}{\big( \Delta^2 + \frac{\kappa^2}{4} \big)^2} \,  G^{2p-2}_{>}(t_{12}) \, \mathcal F_{\lambda}(t_1,t_2) = \frac{\mathsf{A}^2}{\cosh^2 (\mathsf A |t_{12}| + \mathsf B)} \,  \mathcal F_{\lambda}(t_1,t_2) \ .
    \label{eq:middle_equation_derivation_poschl_teller_potential_fermion}
\end{equation}
Let us remark that in this case we have not performed any `inconsistent' approximation.

Putting everything together, and calling $t \equiv t_1 - t_2$, we obtain the second order ODE
\begin{equation}
    \bigg( \! - \frac{\de^2}{\de t^2} - \frac{2 \mathsf{A}^{2}}{\cosh^{2}(\mathsf{A} |t| + \mathsf{B})} - \frac{2(p-1)}{p} \, \mathsf{A} \, \tanh(\mathsf{B}) \, \delta(t) \bigg) f(t) = - \bigg( \Gamma + \frac{\lambda}{2} \bigg)^2 \, f(t) \equiv - \mathsf E^2 \, f(t) \ .
    \label{eq:Poschl-Teller_potential_Lyapunov_QM_problem}
\end{equation}
The solution to the bound state of this ODE gives the (square root of the) energy $\mathsf E$, which in turn is related to the Lyapunov exponent. It is interesting to compare this expression with the corresponding one for large--$p$ SYK, which can be found in~\cite{Maldacena:2016hyu}. We notice that dissipation modifies the ODE in two related ways. On the one hand, it modifies the fermionic two--point function, which results in a non--analytic form of the Pöschl–Teller potential due to the absolute value. On the other hand, it introduces a $\delta$--function contact term, which comes from the Keldysh correlator~\eqref{eq:large-p_bosonic_Keldysh_correlator}, and thus it encodes the effect of quantum fluctuations coming from the bath. It is interesting to notice that the contact term in the potential is attractive, resulting in a bound state with a lower energy and thus increasing the Lyapunov exponent. This effect is crucial to maintain the Lyapunov exponent always positive, and motivates how the bath fluctuations increase scrambling.

The generic solution to~\eqref{eq:Poschl-Teller_potential_Lyapunov_QM_problem} can easily be found, as the Pöschl–Teller has been extensively studied in the literature. The solution (up to normalization) quoted in the main text is
\begin{equation}
    f(t) = e^{- \mathsf{E} ( |t| + \mathsf{B} / \mathsf{A} )}  \Big( \mathsf{E} + \mathsf{A} \tanh \big( \mathsf{A} |t| + \mathsf{B} \big) \Big) \ ,
\end{equation}
namely~\eqref{eq:eigenfunction_Poschl-Teller_potential}. The $\delta$--function imposes a discontinuity in the first derivative of $f(t)$, which results in a quadratic equation for $\mathsf E$. Solving this quadratic equation gives~\eqref{eq:energy_bound_state}, and in turn the Lyapunov exponent~\eqref{eq:Dissipative_Lyapunov_exponent_large_p_analytics}, once all variables have been rewritten in terms of $J_{\rm eff}$ and $\Gamma_{\rm Pur}$.

\end{document}